\documentclass[12pt,draftclsnofoot,onecolumn]{IEEEtran}

\usepackage{amsmath,graphics,amssymb,epsfig,subfigure,color,cite}
\usepackage{array}
\usepackage{multirow}
\usepackage{enumerate}
\usepackage{algorithm}
\usepackage{algpseudocode}
\usepackage{algcompatible}
\usepackage{tablefootnote}
\usepackage[flushleft]{threeparttable}

\algblockdefx{FORP}{ENDFORP}[1]%
  {\textbf{for }#1 \textbf{do in parallel}}%
  {\textbf{end}}
\algblockdefx{FOR}{ENDFOR}[1]%
  {\textbf{for }#1 \textbf{do}}%
  {\textbf{end}}
\algblockdefx{IF}{ENDIF}[1]%
  {\textbf{if }#1 \textbf{then}}%
  {\textbf{end}}
\algblockdefx{PROC}{ENDPROC}[1]
    {\textbf{procedure }#1 {}}
    {\textbf{end procedure}}
  
\algnewcommand\algorithmicprocedure{\textbf{function}}
\algnewcommand\FUNC{\item[\algorithmicprocedure]}%
\algnewcommand\algorithmicendprocedure{\textbf{end function}}
\algnewcommand\ENDFUNC{\item[\algorithmicendprocedure]}%

\usepackage{float}
\usepackage{makecell}
\usepackage[caption=false,font=normalsize,labelfont=sf,textfont=sf]{subfig}

\newtheorem{thm}{Theorem}
\newtheorem{lem}{Lemma}

\newcommand{\bs}{\boldsymbol}

\makeatletter

\newcommand{\vast}{\bBigg@{4.5}}
\newcommand{\Vast}{\bBigg@{7.5}}

\makeatother

\allowdisplaybreaks

\begin{document}
    \title{Communication-Efficient Federated Learning over Capacity-Limited Wireless Networks}
    
	\author{Jaewon Yun, Yongjeong Oh, \IEEEmembership{Student Member,~IEEE}, Yo-Seb Jeon, \IEEEmembership{Member,~IEEE}, \\ and H. Vincent Poor, \IEEEmembership{Life Fellow,~IEEE}
		\thanks{Jaewon Yun, Yongjeong Oh, and Yo-Seb Jeon are with the Department of Electrical Engineering, POSTECH, Pohang, Gyeongbuk 37673, South Korea (e-mail: jaewon.yun@postech.ac.kr; yongjeongoh@postech.ac.kr; yoseb.jeon@postech.ac.kr).}
		\thanks{H. Vincent Poor is with the Department of Electrical and Computer Engineering, Princeton University, Princeton, NJ 08544, USA (e-mail: poor@princeton.edu).}
	}
	\vspace{-2mm}	
	
	\maketitle
	\vspace{-12mm}

	\begin{abstract}
        In this paper, a communication-efficient federated learning (FL) framework is proposed for improving the convergence rate of FL under a limited uplink capacity.
       The central idea of the proposed framework is to transmit the values and positions of the top-$S$ entries of a local model update for uplink transmission. 
       A lossless encoding technique is considered for transmitting the positions of these entries, while a linear transformation followed by the Lloyd-Max scalar quantization is considered for transmitting their values. 
       For an accurate reconstruction of the top-$S$ values, a linear minimum mean squared error method is developed based on the Bussgang decomposition. 
       Moreover, an error feedback strategy is introduced to compensate for both compression and reconstruction errors. 
       The convergence rate of the proposed framework is analyzed for a non-convex loss function with consideration of the compression and reconstruction errors.
       From the analytical result, the key parameters of the proposed framework are optimized for maximizing the convergence rate for the given capacity.
       Simulation results on the MNIST and CIFAR-$10$ datasets demonstrate that the proposed framework outperforms state-of-the-art FL frameworks in terms of classification accuracy under the limited uplink capacity.
	\end{abstract}
	

	\section{Introduction}\label{Sec:Intro}
    Federated learning (FL) is a distributed learning technique for enhancing the privacy of local training datasets generated at edge devices when these datasets are utilized for training a global model on a parameter server (PS) remote from the devices \cite{Konecny:2015,Mcmahan:2017,Konecny:2017}.
    In FL, the PS does not explicitly access the local training datasets, but rather collaborates with the devices for training the global model.
    A typical FL strategy consists of alternating between a local update process at the devices and a global update process at the PS, until the global model converges.
    During the local update process, each device first updates its local model based on the local training dataset, then sends a local model update (i.e., the difference between the local models before and after the update) to the PS. 
    Meanwhile, during the global update process, the PS updates its global model by aggregating the local model updates sent by the devices, then broadcasts the updated global model, or a global model update (i.e., the aggregation of the local model updates), to the devices. 
    By alternating between these two processes, the global model on the PS is trained based on the local training datasets which are kept where they are originally generated. 
    Due to the potential benefits of data privacy, FL has attracted significant attention as a privacy-enhancing solution for machine learning applications in various areas \cite{Niknam:2020,Zhu:2020,Gunduz:2020}.  

    FL often involves two communication processes: (i) {\em uplink transmission}, in which the local model updates are transmitted from the devices to the PS, and (ii) {\em downlink transmission}, in which the updated global model (or global model update) is transmitted from the PS to the devices.
    In practical FL applications, the capacity of the uplink channel from the devices to the PS has a limited capacity. 
    This severely affects the FL convergence rate which decreases as the uplink capacity decreases.
    This problem becomes even more amplified in wireless FL, in which the PS and devices are connected through wireless channels whose capacities are greatly limited due to scarce radio resources. 
    For example, when the capacity of the uplink channel is $20$ Mbps, the typical FL strategy for training the ResNet-$50$ in \cite{ResNet} with $2.6\times10^7$ trainable parameters for $100$ iterations with $10$ devices takes approximately $4.16\times10^4$ s. Such a significant latency may not be acceptable in practical machine learning applications that require faster training in much shorter time.
    
    Communication-efficient FL has been extensively investigated for its potential capability for improving the convergence rate of FL under a limited uplink capacity \cite{Nguyen:21,Mingzhe:21,Hao:22,Jeon:2021}. 
    The most widely adopted approach in this direction consists in the quantization of the local model update prior to the uplink transmission to reduce uplink communication overhead. 
    For example, in \cite{SignSGD}, each entry of the local model update is quantized using a sign quantizer prior to the uplink transmission, which reduces the uplink communication overhead to 1 bit per local update entry. 
    Similarly, in \cite{QSGD}, each entry of the local model update is quantized using a stochastic quantizer for reducing the uplink communication overhead and preserving the statistical property of the local model update as well. 
    FL with scalar quantization was extended in \cite{VeQ:1,VeQ:2,VeQ:3} by simultaneously quantizing multiple entries of the local model update using a vector quantizer. 

    Communication-efficient FL based on sparsification has also been attracting attention owing to its feature of achieving a larger reduction in the communication overhead compared to the quantization-only approach. 
    The fundamental idea of the sparsification is to transmit only the most important information about the local model update. 
    To determine the important information, the magnitudes of the model update entries were considered in \cite{QCS:1,QCS:2,Jeon:2022}, while a low-rank decomposition was employed in \cite{ATOMO,PowerSGD}.
    One of the representative approaches in this direction is to apply quantized compressed sensing (QCS) for compressing the model update after the sparsification. 
    In this approach, the sparsified model update is first projected onto a lower dimensional space and then quantized using a scalar quantizer \cite{SQCS:1,SQCS:2,FedQCS} or a vector quantizer \cite{FedVQCS}. 
    In contrast to the quantization-only approach, the QCS approach provides a better trade-off between the performance and communication overhead of FL in many cases. 
    Nevertheless, the QCS approach is still not suitable for achieving a significant reduction in the communication overhead because the compression level of the dimensionality reduction is theoretically limited to guarantee an accurate reconstruction of the original model update.

        
    Another representative approach for communication-efficient FL via sparsification is a value-position encoding approach which separately encodes the values and positions of the most significant entries of the local or global model update for the uplink transmission \cite{D-DSGD,F. Sattler:2019,TCS}. 
    In these studies, a scalar quantizer is utilized for value encoding, while a lossless encoding technique is utilized for position encoding.
    When the number of the most significant entries is sufficiently small, the value-position encoding approach provides a significant reduction in the communication overhead of FL.
    Thus, it exhibits a considerable advantage over other communication-efficient approaches in supporting FL under a limited uplink capacity. 
    The convergence rate of the FL with the value-position encoding approach, however, is sensitive to the quantizer design and also to the number of the most significant entries to be transmitted.  
    Despite this sensitivity, to the best of our knowledge, no existing study has provided the optimal quantizer design or the optimal number of the most significant entries for maximizing the  convergence rate of the FL with the value-position encoding approach. 

    In this paper, we propose a new communication-efficient FL framework via sparsification, dubbed as {\em FedSpar}. 
    In this framework, following the idea of the value-position encoding approach, we develop compression and reconstruction strategies for the uplink transmission in FL. 
    We also analyze the convergence rate of the proposed FL framework and determine the optimal number of the most significant entries to be transmitted for maximizing the convergence rate.
    The major contributions of this paper include the following:
    
    \begin{itemize}
        \item 
        We develop a new compression strategy for uplink transmission in FL, which reduces the compression error of the value-position encoding. 
        The basic idea of our strategy is to encode the values and positions of the top-$S$ entries of the local model update in terms of their magnitudes. 
        A similar compression strategy was introduced in \cite{D-DSGD,F. Sattler:2019} in which the values of the top-$S$ entries are encoded by a heuristic scalar quantizer.
        Unlike in these works, in our strategy, we perform linear transformation before quantizing the top-$S$ values, so that the transformed values can be modeled as Gaussian random variables.
        We then design the optimal scalar quantizer for the corresponding Gaussian distribution in order to minimize the compression error of the value encoding.
        
        \item 
       We develop a new reconstruction strategy for uplink reception in FL, which minimizes the reconstruction error of the value-position decoding.
        In general, minimizing the reconstruction error in decoding the top-$S$ values is challenging due to the lack of knowledge of their distributions. 
        In our reconstruction strategy, we address this challenge by deriving the linear minimum-mean-squared-error (LMMSE) estimator for the top-$S$ value decoding based on the Gaussian modeling and the Bussgang decomposition in \cite{J. J. Bussgang:1952}.

        \item
       We analyze the convergence rate of the proposed FedSpar framework for a non-convex loss function. 
        When lossy compression and reconstruction are applied in both uplink and downlink communications, it is generally challenging to analyze the convergence rate of FL due to complicated correlations among different types of the errors. 
        This challenge is addressed in FedSpar by employing the error feedback strategy in both uplink and downlink communications, which compensates for joint effects of the different types of the errors.
        Leveraging this advantage, we prove that FedSpar converges to a stationary point of a non-convex loss function under consideration of the compression and reconstruction errors in the uplink and downlink communications.

        
        \item
        We optimize the key parameters of the compression strategy of FedSpar, which are (i) a sparsification level (i.e., the number of entries chosen during the sparsification) and (ii) a quantization level for the value encoding.
        Although the top-$S$ value encoding using a scalar quantizer was considered in \cite{D-DSGD,F. Sattler:2019}, none of these works have explored the optimization of the sparsification level $S$ and the quantization level $Q$.
        In this work, we characterize the optimal sparsification and quantization levels for maximizing the convergence rate of FedSpar under the constraints of the uplink capacity.

        
        \item
        Using simulations, we demonstrate the superiority of FedSpar for an image classification task using the MNIST dataset \cite{MNIST} and the CIFAR-$10$ dataset \cite{CIFAR-10}. 
        Our results show that FedSpar outperforms the state-of-the-art FL frameworks in terms of classification accuracy under the constraint of uplink capacity. 
        Meanwhile, for the MNIST dataset, FedSpar with $0.4$-bit overhead per entry for the uplink transmission yields only a $1$\% decrease in classification accuracy compared to the vanilla FL framework with lossless uplink transmission.
        We also demonstrate the effectiveness of the error feedback strategy and the parameter optimization in improving the classification accuracy of FedSpar.
    \end{itemize}
    
    {\em Notation:} Upper-case and lower-case boldface letters denote matrices and column vectors, respectively. 
    $\mathbb{E[\cdot]}$ is the statistical expectation, and $(\cdot)^{\sf T}$ is the transpose. $|\mathcal{A}|$ is the cardinality of set $\mathcal{A}$. 
    $(\boldsymbol{a})_i$ represents the $i$-th element of vector $\boldsymbol{a}$. 
    $\|\boldsymbol{a}\| = \sqrt{\boldsymbol{a}^{\sf T}\boldsymbol{a}}$ is the Euclidean norm of a real vector $\boldsymbol{a}$. 
    $\mathcal{N}({\bs \mu},{\bs R})$ represents the distribution of a Gaussian random distribution with mean vector $\boldsymbol{\mu}$ and covariance matrix $\boldsymbol{R}$. 
    $\boldsymbol{0}_n$ and $\boldsymbol{1}_n$ are $n$-dimensional vectors whose elements are zero and one, respectively. 
    $\boldsymbol{I}_N$ is an $N$ by $N$ identity matrix.
    
    \section{System Model}
    Consider a wireless FL scenario in which a PS trains a global model (e.g., deep neural network) by collaborating with a set $\mathcal{K}$ of $K$ wireless devices. 
    The global model on the PS is represented by a global parameter vector $\boldsymbol{w}\in \mathbb{R}^{{N}}$, where ${N}$ is the number of global model parameters.
    The goal of the PS is to minimize a \textit{global} loss function defined as
	\begin{align}\label{eq:global_loss0}
    	F(\boldsymbol{w}) \triangleq \frac{1}{|\mathcal{D}|} \sum_{\boldsymbol{u}\in \mathcal{D}} f(\boldsymbol{w};\boldsymbol{u}),
	\end{align}
	by training the global parameter vector $\boldsymbol{w}$, where $\mathcal{D}$ is a training dataset, and $f(\boldsymbol{w};\boldsymbol{u})$ is a loss function computed for $\boldsymbol{w}$ with a training data sample $\boldsymbol{u}$.
	Without an FL scenario, a typical approach for training the global parameter vector consists in employing an iterative gradient-based algorithm, in which the global parameter vector $\boldsymbol{w}^{(t)}$ at iteration $t$ is updated based on a {\em true} gradient vector defined as
	\begin{align}\label{eq:true_grad}
	    \nabla F (\boldsymbol{w}^{(t)}) 
	    \triangleq \frac{1}{|\mathcal{D}|}  \sum_{\boldsymbol{u}\in \mathcal{D}} \nabla f(\boldsymbol{w}^{(t)};\boldsymbol{u}),
	\end{align}
    for all $t\in\{1,\ldots,T\}$, where $\nabla$ is a gradient operator, and $T$ is the  number of the iterations.
	For example, if the gradient descent algorithm is employed at the PS, the update rule at iteration $t$ would be given by  
	\begin{align}\label{eq:global_update}
	    \boldsymbol{w}^{(t+1)} \leftarrow \boldsymbol{w}^{(t)} - \eta^{(t)} \nabla F (\boldsymbol{w}^{(t)}), ~~\forall t\in\{1,\ldots,T\},
	\end{align}
	where $\eta^{(t)} >0$ is a learning rate at iteration $t$. 

	In the FL scenario, it is common to assume that the training dataset $\mathcal{D}$ is distributed over the wireless devices, whereas the PS is not allowed to explicitly access the training dataset. 
	Let $\mathcal{D}_k$ be a {\em local} training dataset available at device $k \in \mathcal{K}$ such that $\mathcal{D} = \cup_k \mathcal{D}_k$.
	Then the global loss function in \eqref{eq:global_loss0} is rewritten as 
	\begin{align}\label{eq:global_loss}
    	F(\boldsymbol{w}) 
    	= \frac{1}{\sum_{j=1}^K|\mathcal{D}_j|} \sum_{k=1}^K  |\mathcal{D}_k|  F_k(\boldsymbol{w}),
	\end{align}
	where $F_k({\bf w})$ is a \textit{local} loss function computed by device $k$, defined as
	\begin{align}
	    F_k(\boldsymbol{w}) \triangleq \frac{1}{|\mathcal{D}_k|} \sum_{\boldsymbol{u}\in\mathcal{D}_k} f(\boldsymbol{w};\boldsymbol{u}).
	\end{align}	
	In this scenario, the PS collaborates with the wireless devices for minimizing the global loss function in \eqref{eq:global_loss} by training the global parameter vector. 
    To this end, a typical FL framework involves alternating between a local update process at wireless devices and a global update process at the PS.
    Details of each process are described below.
    \begin{itemize} 
        \item {\bf Local update process:}
        Let $\mathcal{K}_t \subset \mathcal{K}$ be a subset of the devices participates in the learning process at iteration $t$ (i.e., partial device participation) such that $|\mathcal{K}_t|=M$.
        Suppose that the devices in $\mathcal{K}_t$ successfully receive the global parameter vector $\boldsymbol{w}^{(t)}$ sent by the PS.
        In the local update process at iteration $t$, device $k \in \mathcal{K}_t$ {\em locally} updates $\boldsymbol{w}^{(t)}$ based on its local training dataset $\mathcal{D}_k$.
        Assume that the devices employ a mini-batch gradient descent algorithm with $E\geq 1$ local iterations for the local update process.
    	Then, by setting $\boldsymbol{w}^{(t,1)}_k = \boldsymbol{w}^{(t)}$, the local parameter vector computed by device $k\in \mathcal{K}_t$ at local iteration $e$ is expressed as
    	\begin{align}
            \boldsymbol{w}^{(t,e+1)}_k = \boldsymbol{w}^{(t,e)}_k - \gamma^{(t)}\nabla F_{k}^{(t,e)}\big(\boldsymbol{w}^{(t,e)}_k \big),
    	\end{align}
    	for $e \in \{1,\ldots,E\}$, where $\gamma^{(t)}$ is a local learning rate, and $\nabla F_{k}^{(t,e)}\big(\boldsymbol{w}^{(t,e)}_k\big)$ is a gradient vector computed by  device $k\in \mathcal{K}_t$ at local iteration $e$, defined as
    	\begin{align}\label{eq:local_grad0}
    	    \nabla F_{k}^{(t,e)}\big(\boldsymbol{w}^{(t,e)}_k\big) \triangleq \frac{1}{|\mathcal{D}_k^{(t,e)}|} \sum_{\boldsymbol{u}\in\mathcal{D}_k^{(t,e)}} \nabla f(\boldsymbol{w}^{(t,e)}_k;\boldsymbol{u}),
    	\end{align}
    	where $\mathcal{D}_k^{(t,e)}$ is a mini-batch randomly drawn from $\mathcal{D}_k$ at local iteration $e$.
        Following the local update process, device $k\in \mathcal{K}_t$ sends a {\em local model update} to the PS, defined as the difference between the local parameter vectors before and after the local update process:
    	\begin{align}\label{eq:local_model_update}
            \boldsymbol{g}^{(t)}_k &\triangleq {\sf LocalUpdate}(\boldsymbol{w}^{(t)},\mathcal{D}_k)= \frac{1}{\gamma^{(t)}E}(\boldsymbol{w}^{(t,1)}_k-\boldsymbol{w}^{(t,E+1)}_k) \in \mathbb{R}^{{N}}.
    	\end{align}
        The transmission from the participating devices to the PS is referred to as the {\em uplink} transmission.

        \item {\bf Global update process:}
        Suppose that the PS successfully receives the local model updates $\{\boldsymbol{g}_k^{(t)}\}_{k \in \mathcal{K}_t}$ sent by $M$ devices at iteration $t$.
        Thus, in the corresponding global update process, the PS computes a {\em global model update} by aggregating $M$ local model updates as follows: 
        \begin{align}\label{eq:global_model_update_true}
    	    \boldsymbol{g}_{\rm PS}^{(t)} = \sum_{k \in \mathcal{K}_t} \rho_k\boldsymbol{g}_{k}^{(t)},
        \end{align}
        where $\rho_k \triangleq \frac{\sum_{e=1}^{E}|\mathcal{D}_k^{(t,e)}|}{\sum_{j \in \mathcal{K}_t} \sum_{e=1}^{E}|\mathcal{D}_j^{(t,e)}|}$.
        The PS then updates the global parameter vector based on the global model update $\boldsymbol{g}_{\rm PS}^{(t)}$ as follows:
        \begin{align}\label{eq:global parameter update}
    	    \boldsymbol{w}^{(t+1)} \leftarrow {\boldsymbol w}^{(t)} -   \eta^{(t)} {\boldsymbol g}_{\rm PS}^{(t)}.
        \end{align}
        Following the global update process, the PS sends the updated global parameter vector $\boldsymbol{w}^{(t+1)}$ to the devices $k\in\mathcal{K}_{t+1}$. 
        The transmission from the PS to the participating devices is referred to as the {\em downlink} transmission.
	\end{itemize}
	\vspace{1mm}
	
	\begin{figure*}
		\centering 
		{\epsfig{file=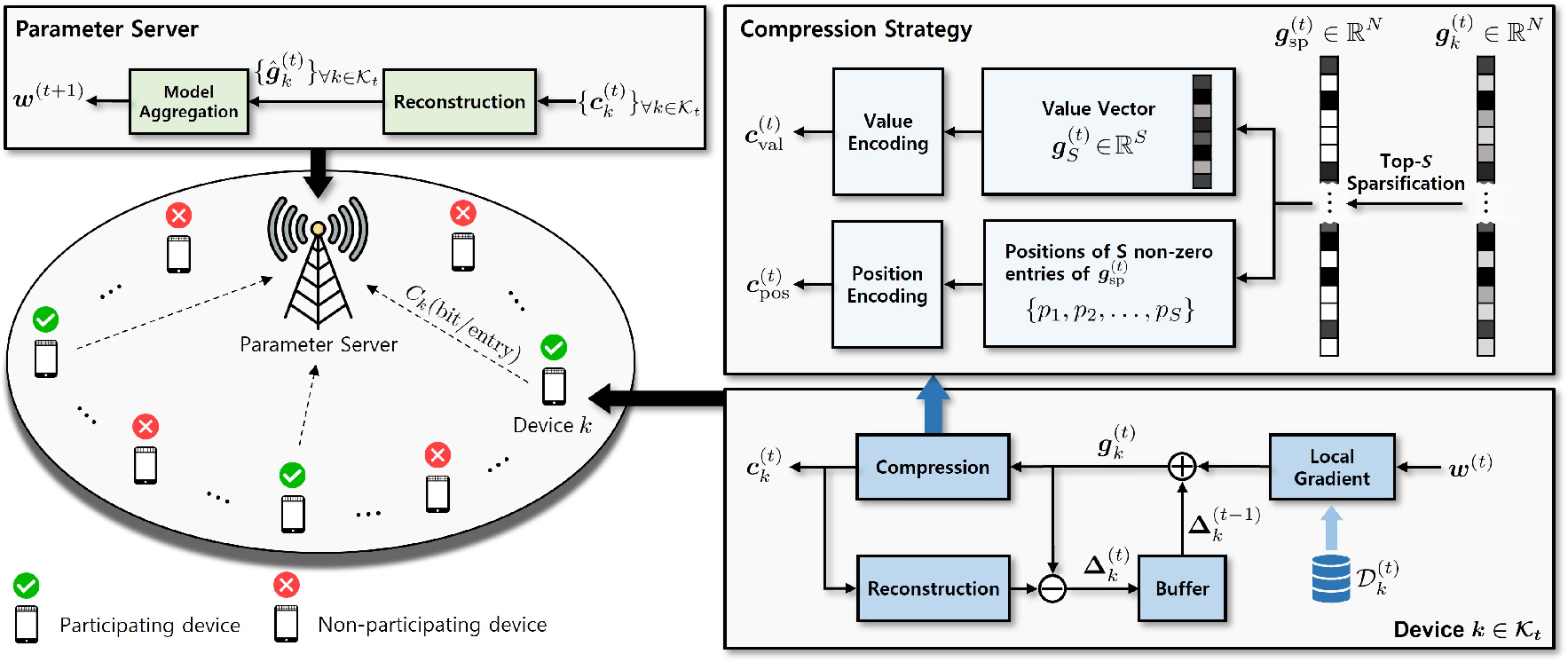, width=16.5cm}}
		\caption{Illustration of the proposed federated learning (FL) framework where the PS and $K$ devices are connected via wireless channels with limited capacities.} \vspace{-5mm}
		\label{fig:FedSpar}
	\end{figure*}
	    
    In this work, we focus on a wireless FL scenario, in which the PS and the wireless devices are connected via wireless channels with limited capacities, as illustrated in Fig.~\ref{fig:FedSpar}.   
    To quantify the capacity of the uplink channel, we assume that the maximum number of bits that can be reliably transmitted from device $k$ to the PS per iteration is given by $C_kN$ with $C_k>0$ for all $k\in\mathcal{K}$. 
    By the definition, the constant $C_{k}$ (measured in bits per entry) depends on the uplink channel capacity (measured in bits per second) and transmission time (measured in seconds).
   If $C_k \ll 32$, it becomes essential to compress the information transmitted by the devices for the given uplink capacity. 
    Concurrently, it is critical to minimize the decrease in the convergence rate of FL  caused by uplink compression and reconstruction errors.

    \section{Federated Learning via Sparsification (FedSpar)}\label{Sec:FedSpar}
    In this section, we propose a novel FL framework, dubbed as {\em FedSpar}, which involves sparsifying the local model updates for improving the communication efficiency of FL under a limited uplink capacity. 
    The major components of FedSpar are (i) a compression strategy for uplink transmission by the participating devices, (ii) a reconstruction strategy for uplink reception by the PS, and (iii) an error feedback strategy to compensate for both compression and reconstruction errors. 
    In the following, we first describe the details of each strategy, and then summarize the overall procedure of FedSpar. 
    	
    \subsection{Compression Strategy of FedSpar}\label{Sec:Compression}
    The key idea of our compression strategy is to transmit only the most significant entries of the local update, achieved by separately encoding their values and positions. Specifically, a {\em lossy} encoding is applied for compressing the values of these entries, whereas their positions are represented by applying a {\em lossless} encoding.
    On the basis of this idea, our compression strategy sequentially performs (i) sparsification, (ii) value encoding, and (iii) position encoding. 
    This strategy can be leveraged for compressing the local model update ${\boldsymbol g}_k^{(t)}$ at device $k \in \mathcal{K}_t$.
    For the sake of simplicity, a simplified notation ${\boldsymbol g}^{(t)}$ without device index $k$ 
    shall be used in the remainder of this section.
    
    \subsubsection{Sparsification}
    In the sparsification step, the $S$ largest entries of the local model update are selected for transmission after encoding, where $S \ll N$ is a sparsification level. The optimization of this level will be discussed further in Sec.~\ref{Sec:Parameter}.
    During sparsification at iteration $t$, the sparsified local model update, namely ${\bs g}_{\rm sp}^{(t)} \in \mathbb{R}^{N}$, is constructed by setting the magnitude to zero for all entries of ${\bs g}^{(t)}$, except the corresponding $S$ largest entries. 
    It follows that the sparsified local model update ${\bs g}_{\rm sp}^{(t)}$ is an $S$-sparse vector, with $S$ non-zero entries whose values and positions are compressed by the subsequent value and position encoding steps.

    \subsubsection{Value Encoding}
   In the value encoding step, a linear transformation, followed by scalar quantization, are used to encode the values of the $S$ non-zero entries of the sparsified local model update ${\bs g}_{\rm sp}^{(t)}$.
    To this end, we first define a value vector ${\bs g}_{S}^{(t)} \in \mathbb{R}^S$ whose entries correspond to the $S$ non-zero entries of ${\bs g}_{\rm sp}^{(t)}$, i.e., ${\bs g}_S^{(t)} = \big[g_{{\rm sp},p_1}^{(t)},\cdots, g_{{\rm sp},p_S}^{(t)}\big]^{\sf T}$, where $g_{{\rm sp},p_i}^{(t)}$ is the $i$-th non-zero entry of ${\bs g}_{\rm sp}^{(t)}$.
    Then, we compress the value vector by sequentially applying (i) normalization, (ii) linear transformation, and (iii) quantization. 
    
    \begin{itemize}
        \item {\bf Normalization:}
        In this process, we normalize the value vector ${\bs g}_{S}^{(t)}$ so that its entries have zero mean and unit variance. 
        For this, we first compute the sample mean and variance of the entries of ${\bs g}_{S}^{(t)}$, given by
        \begin{align}
            {\mu}_{S}^{(t)} &= {\bs 1}_S^{\sf T} {\boldsymbol g}_{S}^{(t)} /S,\label{eq:mean} \\
            {\nu}_{S}^{(t)} &= \| {\boldsymbol g}_{S}^{(t)}-{\mu}_{S}^{(t)}{\bs 1}_S \|^2/S = \| {\boldsymbol g}_{S}^{(t)}\|^2/S - \big({\mu}_{S}^{(t)} \big)^2,
            \label{eq:variance}
        \end{align}
        respectively, where ${\bs 1}_S = [1,\ldots,1]^{\sf T}$ is an $S$-dimensional all-one vector.
        Utilizing the above mean and variance, we normalize the value vector as follows:
        \begin{align}\label{eq:normalize}
    	    {\bs v}^{(t)} =  ({\bs g}_{S}^{(t)} - {\mu}_{S}^{(t)}{\bs 1}_S)/{\sqrt {{\nu}_{S}^{(t)}}}.
    	\end{align}
    	Then the sample mean and variance of the entries for the normalized value vector ${\boldsymbol v}^{(t)}$ become zero and one, respectively. 

    	\item {\bf Linear transformation:}
    	In this process, we transform the normalized value vector ${\bs v}^{(t)}$ using a random orthogonal matrix ${\bs U} \in \mathbb{R}^{S \times S}$ such that ${\bs U}{\bs U}^{\sf T} = {\bs I}_S$.
    	Let ${\boldsymbol x}^{(t)} \in \mathbb{R}^{S}$ be the transformed value vector at iteration $t$, determined by
        \begin{align}\label{eq:transform}
            {\boldsymbol x}^{(t)} = \big[ x_{1}^{(t)}, \cdots,  x_{S}^{(t)} \big]^{\sf T} = {\boldsymbol U}{\boldsymbol v}^{(t)}.
        \end{align}    	
        In particular, we use a Haar-distributed orthogonal matrix ${\bs U}$, generated by performing the Gram-Schmidt procedure on a random matrix whose entries are independent standard Gaussian random variables. 
    	In this case, any $\mathcal{O}(S)$ entries of the Haar-distributed orthogonal matrix asymptotically behave like independent Gaussian random variables with zero mean and variance $1/S$.
    	This implies that for large $S$, the entries of the transformed value vector ${\boldsymbol x}^{(t)}$ behave like independent standard Gaussian random variables (i.e., $x_s^{(t)} \sim \mathcal{N}(0,1)$, $\forall s$) by the central limit theorem and the normalization in \eqref{eq:normalize}.
        
        \item {\bf Quantization:} 
        In this process, we quantize the transformed value vector ${\boldsymbol x}^{(t)}$ using a $Q$-level scalar quantizer.
        Let ${\sf Q}: \mathbb{R} \rightarrow \mathcal{Q}\triangleq \{q_1,\ldots,q_{Q}\}$ be a quantization function of the scalar quantizer, which satisfies that ${\sf Q}(x) = q_i$ if $x \in (\tau_{i-1},\tau_i]$, where $q_i$ is the $i$-th quantizer output, and $\tau_i$ is the $i$-th quantizer threshold such that $\tau_0 <  \ldots < \tau_{Q}$ with $\tau_0 = -\infty$ and $\tau_{Q} = \infty$. 
    	Then, the quantized value vector is obtained as ${\boldsymbol q}^{(t)} = \big[ {\sf Q}(x_{1}^{(t)}), \cdots,  {\sf Q}(x_{S}^{(t)})\big]^{\sf T}$.
    	The design of the scalar quantizer for a given quantization level $Q$ is critical for reducing the reconstruction error of the model update at a receiver side. 
    	Note that for large $S$, the distribution of the quantizer's input is given by $x_s^{(t)} \sim \mathcal{N}(0,1)$, $\forall s$, as discussed above.
    	Motivated by this, we harness the Lloyd--Max quantizer in \cite{S. P. Lloyd:1982} optimized for the Gaussian distribution with zero mean and unit variance. 
    	Notably, the Lloyd--Max quantizer minimizes the mean-squared-error (MSE) of the quantizer output for the given quantization level. 
     The devices and the PS can generate the same set of the Lloyd-Max quantizers by running the Lloyd-Max algorithm \cite{S. P. Lloyd:1982} for each quantization level $Q\in\{2,3,\dots,Q_{\rm max}\}$ with respect to an input distribution $\mathcal{N}(0,1)$.
    \end{itemize}
    The aforementioned step (i.e., value encoding) necessitates the transmission of the mean $\mu_S^{(t)}$, variance ${\nu}_S^{(t)}$, and quantized value vector ${\boldsymbol q}^{(t)}$.
    Therefore, the bit overhead for transmitting the value vector ${\boldsymbol g}_S^{(t)}$ using the value encoding step is given by $B_{\rm  val} = S\log_2Q + 64$.

    \subsubsection{Position Encoding}
    In this step, the positions of the $S$ non-zero entries of the sparsified model update ${\boldsymbol g}_{\rm sp}^{(t)}$ are encoded using a lossless coding technique. 
    A trivial example of such a technique is to send the index of the combination, corresponding to the positions of the $S$ non-zero entries of ${\boldsymbol g}_{\rm sp}^{(t)}$, among all possible combinations of $S$ non-zero positions taken from $N$ positions.
    Let $p_i$ be the position of the $i$-th non-zero entry of ${\bs g}_{\rm sp}^{(t)}$ and let $\mathcal{P}({\bs g}_{\rm sp}^{(t)}) = \{p_1,p_2,\ldots,p_S\}$ be the set of the positions of the $S$ non-zero entries of ${\bs g}_{\rm sp}^{(t)}$. 
    Then the set $\mathcal{P}({\bs g}_{\rm sp}^{(t)})$ can be readily mapped to a number $i \in \{1,\ldots,\binom{N}{S}\}$ by using the ranking algorithm in \cite{Lexicographic}.
    With this technique, the bit overhead for transmitting the position information is  given by $B_{\rm pos} = \log_2 \binom{N}{S}$.
    We can further reduce the above bit overhead by employing advanced source coding techniques such as the Golomb and Huffman encoding. Nonetheless, it is worth noting that the bit overhead of these techniques is not deterministic. 
    To avoid this uncertainty, we only consider the simple combination-based technique as a worst case example.

    \begin{algorithm}[t]
    	\caption{Compression Strategy of FedSpar}\label{alg:Compress}
    	{\small
    	{\begin{algorithmic}[1]
                \REQUIRE The local model update ${\bs g}^{(t)}$, the sparsification level $S$, and the quantization level $Q$
                \ENSURE ${\bs c}^{(t)}$
    	    \PROC {${\sf Compress} \big({\bs g}^{(t)}; S, Q \big)$}
    		    \STATE Construct ${\bs g}_{\rm sp}^{(t)}$ by setting all but the $S$ largest entries of ${\bs g}^{(t)}$ to zero
    		    \STATE ${\bs g}_S^{(t)} = \big[g_{{\rm sp},p_1}^{(t)},\cdots, g_{{\rm sp},p_S}^{(t)}\big]^{\sf T}$
    		    \STATE ${\mu}_{S}^{(t)} = {\bs 1}_S^{\sf T} {\boldsymbol g}_{S}^{(t)} /S$
    		    \STATE ${\nu}_{S}^{(t)} = \| {\boldsymbol g}_{S}^{(t)}\|^2/S - \big({\mu}_{S}^{(t)} \big)^2$
    		    \STATE ${\bs v}^{(t)} =  ({\bs g}_{S}^{(t)} - {\mu}_{S}^{(t)}{\bs 1}_S)/{\sqrt {{\nu}_{S}^{(t)}}}$
    		    \STATE ${\boldsymbol x}^{(t)} ={\boldsymbol U}{\boldsymbol v}^{(t)}$
    		    \STATE ${\boldsymbol q}^{(t)} = \big[ {\sf Q}(x_{1}^{(t)}), \cdots,  {\sf Q}(x_{S}^{(t)})\big]^{\sf T}$
    		    \STATE Determine a bit sequence ${\bs c}_{\rm val}$ representing ${\mu}_{S}^{(t)}$, ${\nu}_{S}^{(t)}$, and ${\boldsymbol q}^{(t)}$
    		    \STATE Determine a bit sequence ${\bs c}_{\rm pos}$ representing the index of $\mathcal{P}({\bs g}_{\rm sp}^{(t)})$
                \STATE \textbf{return} ${\bs c}^{(t)} = [{\bs c}_{\rm val}^{\sf T}, {\bs c}_{\rm pos}^{\sf T}]^{\sf T}$
            \ENDPROC
    	\end{algorithmic}}}
    \end{algorithm}
    \vspace{2mm}
    
    In {\bf Procedure~\ref{alg:Compress}}, we summarize the aforementioned compression strategy of  FedSpar, denoted by ${\sf Compress}({\bs g}^{(t)};S,Q)$, which is applied to a model update ${\bs g}^{(t)}$ with the sparsification level $S$ and the quantization level $Q$.
    Following this strategy, each participating device transmits a bit sequence representing the outputs of the value and position encoding during the uplink transmission. The total bit overhead of our compression strategy is given by
    \begin{align}\label{eq:total_bit}
        B_{\rm tot} = B_{\rm val} + B_{\rm pos} = S\log_2Q + 64 + \log_2 \binom{N}{S}.
    \end{align}	
    Assume that device $k\in\mathcal{K}_t$ adjusts the sparsification level $S$ and the quantization level $Q$ so that the corresponding bit overhead satisfies $B_{\rm tot} \leq C_k N$ for the uplink transmission. 
    With this assumption, we can guarantee error-free transmission of the bit sequence even for noisy uplink channel because we already define $C_k$ to capture the limited capacities of these channels, as discussed in Sec.~II.
    The optimization of $S$ and $Q$ will be discussed in Sec.~\ref{Sec:Parameter}.
    The bit sequence representing the encoding output of the local model update ${\boldsymbol g}_k^{(t)}$ can be denoted by ${\boldsymbol c}_k^{(t)} \in \{0,1\}^{C_k N}$.
    
    \begin{figure*}
		\centering 
		\subfigure[$T=1$]
		{\epsfig{file=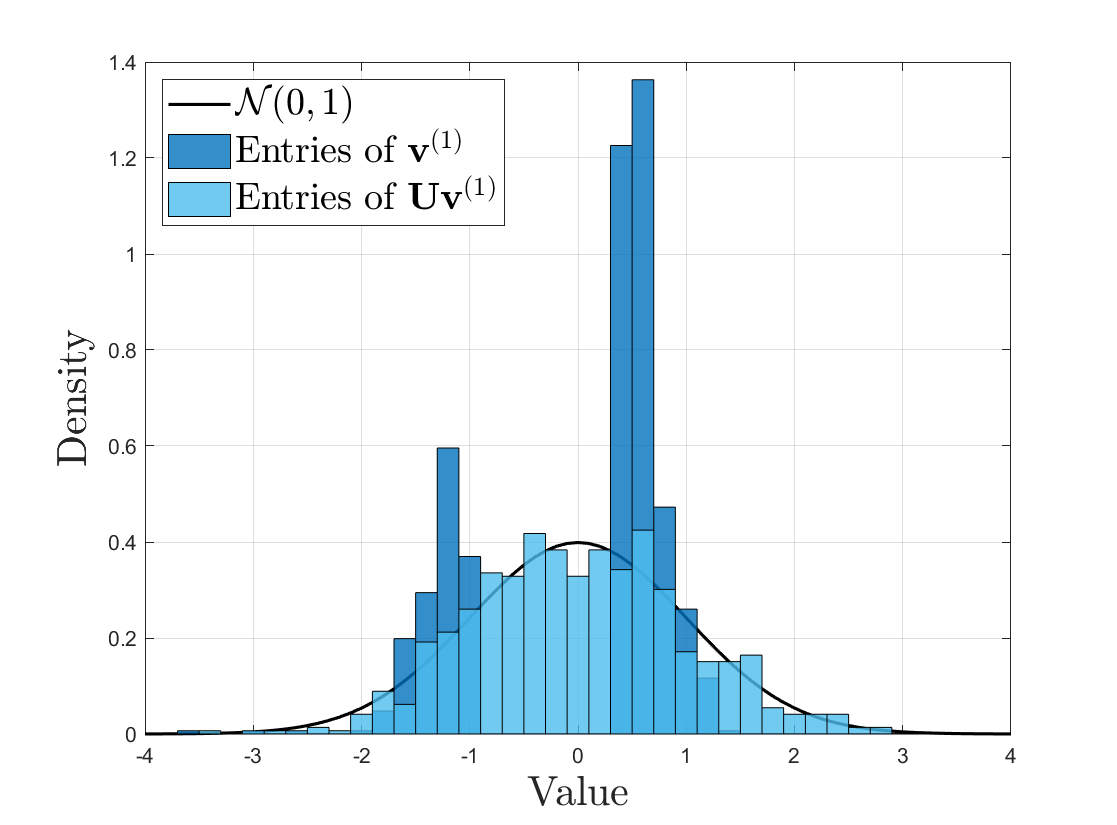, width=5.41cm}}
		\subfigure[$T=50$]
		{\epsfig{file=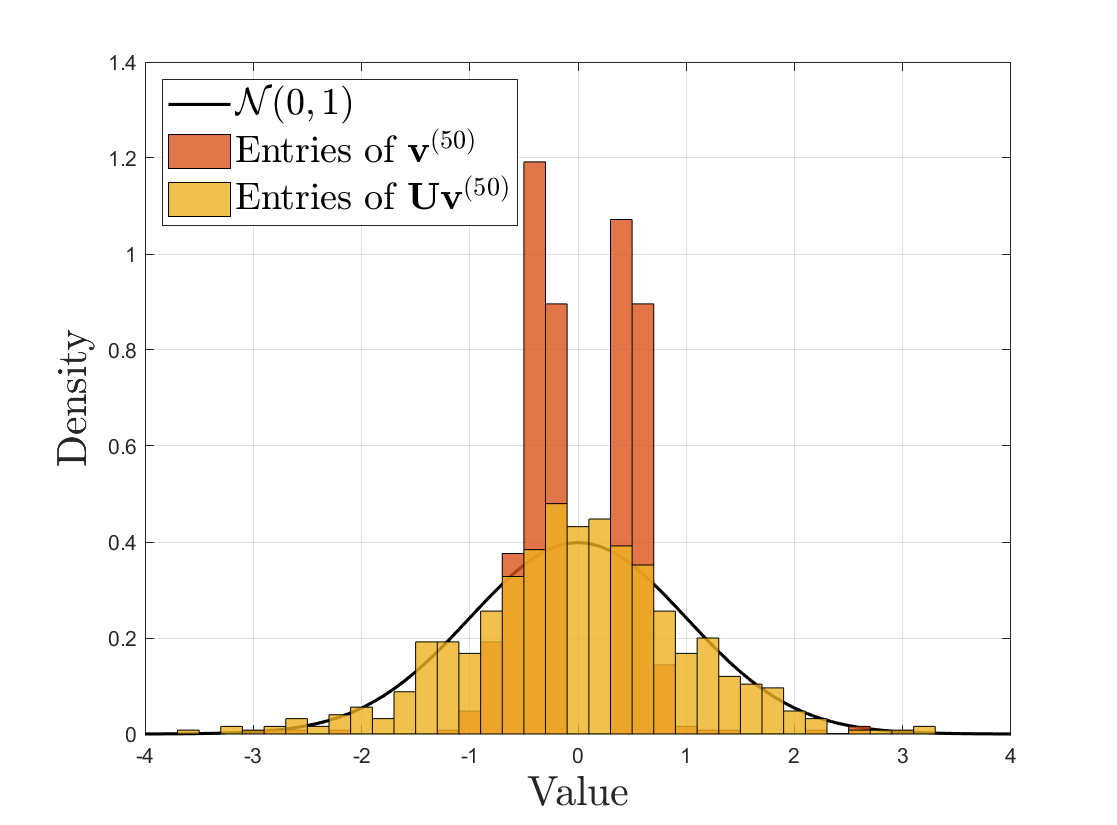, width=5.41cm}}
		\subfigure[$T=100$]
		{\epsfig{file=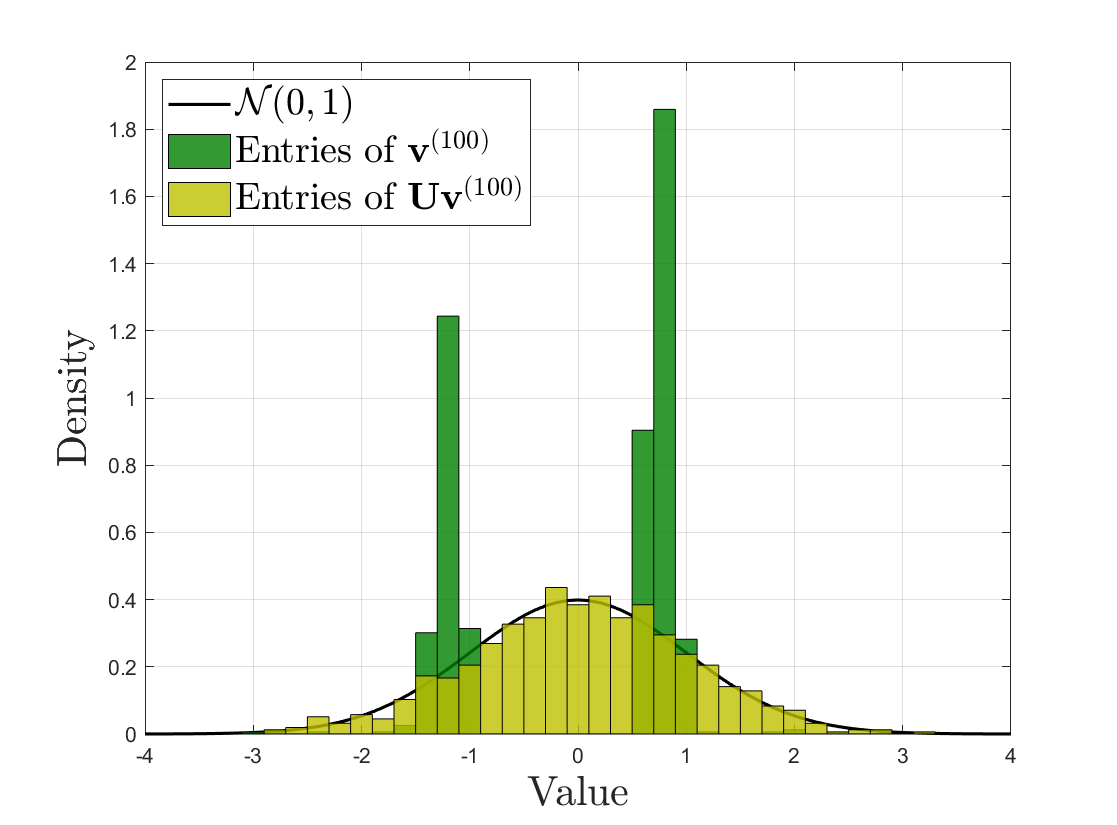, width=5.41cm}}
		\vspace{-2mm}
		\caption{Comparison of the probability density function (PDF) of the standard Gaussian random variable with the histograms of the entries of the transformed value vector and the value vector randomly sampled during the uplink transmission after $T$ iterations.} 
        \vspace{-5mm}
		\label{fig:gaussian}
	\end{figure*}
    \vspace{2mm}
    
    {\bf Numerical example (Gaussian modeling):}
    In this numerical example, we further justify the Gaussian modeling of the transformed value vector in \eqref{eq:transform}. 
    To this end, we compare the probability density function (PDF) of the standard Gaussian random variable (i.e., $\mathcal{N}(0,1)$) to the histograms of the entries of the transformed value vector ${\bs U}{\bs v}^{(T)}$ and the normalized value vector ${\bs v}^{(T)}$ randomly sampled during the uplink transmission in $T$ iteration $t$, as shown in Fig.~\ref{fig:gaussian}.
    In this simulation, we consider an image classification task using the MNIST dataset described in Sec.~\ref{Sec:Simul} when $C_k = 0.4$ bits/entry for all $k\in\mathcal{K}$.
    Fig.~\ref{fig:gaussian} shows that the entries of the value vector after the linear transformation can be well modeled as the standard Gaussian random variables. 
    This result verifies the tightness of the Gaussian modeling of the transformed value vector (i.e., $x_s^{(t)} \sim \mathcal{N}(0,1)$, $\forall s$)  discussed above. 
    Meanwhile, Fig.~\ref{fig:gaussian} indicates that the distribution of the entries of the value vector, without the linear transformation, significantly differs from that of the Gaussian random variables. 
    This implies that the linear transformation using the Haar-distributed orthogonal matrix is crucial for attaining the Gaussian-like entries, which can facilitate the optimal design of the scalar quantizer.

    \subsection{Reconstruction Strategy of FedSpar}\label{Sec:Reconstruction}
    The aim of our reconstruction strategy at the PS is to facilitate the accurate reconstruction of the sparsified local model update ${\boldsymbol g}_{\rm sp}^{(t)}$ from the received bit sequence when employing the compression strategy in Sec.~\ref{Sec:Compression}.
    Fortunately, the positions of the non-zero entries of ${\boldsymbol g}_{\rm sp}^{(t)}$ can be perfectly reconstructed from the position-encoded bits by using the unranking algorithm in \cite{Lexicographic}. 
    Therefore, we focus only on developing a technique for accurately reconstructing the values of the non-zero entries of ${\boldsymbol g}_{\rm sp}^{(t)}$ (i.e., the value vector ${\boldsymbol g}_S^{(t)}$).
    We recall that, during the value encoding, the value vector after normalization is linearly transformed by the matrix ${\boldsymbol U}$, and non-linearly distorted by the quantization function ${\sf Q}(\cdot)$. 
    To promote the accurate reconstruction under these two effects, we first find the LMMSE estimate of the transformed value vector ${\bs x}^{(t)}$ based on the Bussgang decomposition in \cite{J. J. Bussgang:1952}. Then, we reconstruct the value vector ${\bs g}_S^{(t)}$ from the LMMSE estimate. 
        
    Our reconstruction strategy is motivated by the fact that the entries of the transformed value vector ${\boldsymbol x}^{(t)} = {\bs U}{\bs v}^{(t)}$ can be modeled as independent standard Gaussian random variables (i.e., ${\boldsymbol x}^{(t)}  \sim \mathcal{N}({\bf 0}_{S},{\bf I}_{S})$), which has been already justified in Sec.~\ref{Sec:Compression} for large $S$.
	Using this, we apply the Bussgang decomposition to the quantized value vector ${\boldsymbol q}^{(t)}$ which yields \cite{FedQCS} 
    \begin{align}\label{eq:Bussgang}
        {\boldsymbol q}^{(t)} = {\sf Q}({\boldsymbol x}^{(t)})  = \gamma_{Q}  {\boldsymbol x}^{(t)}  + {\boldsymbol d}^{(t)}, 
    \end{align}	
    where ${\bs d}^{(t)}$ is an additive distortion uncorrelated with ${\bs x}^{(t)}$, and $\gamma_{Q}$ and $\psi_Q$ are quantizer-dependent constants defined as
    \begin{align}\label{eq:gamma_Q}
        \gamma_{Q} = \sum_{i=1}^{Q} \frac{q_i}{\sqrt{2\pi}} \left\{ \exp\left(-    \frac{\tau_{i-1}^2}{2}\right) - \exp\left(- \frac{\tau_{i}^2}{2}\right) \right\},
    \end{align}	
    and     
    \begin{align}\label{eq:psi_Q}
        \psi_Q = \sum_{i=1}^{Q} q_i^2 \int_{\tau_{i-1}}^{\tau_i} \frac{1}{\sqrt{2\pi}} e^{-\frac{u^2}{2}} {\rm d}u,
    \end{align}
    respectively.
    In \cite{FedQCS}, it is also shown that $\mathbb{E}[{\bs d}^{(t)}] = {\bs 0}_S$ and ${\bs  R}_{{\bs d}^{(t)}} \triangleq \mathbb{E}\big[{\bs d}^{(t)}({\bs d}^{(t)})^{\sf T}\big] = \big(\psi_Q - \gamma_{Q}^2\big) {\bs I}_{S}$.
    The LMMSE estimate of ${\bs x}^{(t)}$, which minimizes the MSE between ${\bs x}^{(t)}$ and its linear estimate from the observation ${\bs q}^{(t)}$ in \eqref{eq:Bussgang}, can thus be characterized.
    Because ${\boldsymbol x}^{(t)}$ has zero mean for large $S$, the LMMSE estimate of ${\bs x}^{(t)}$ is expressed as $\hat{\bs x}^{(t)}  = {\bs F}^{(t)} {\bs q}^{(t)}$, where
	\begin{align}\label{eq:LMMSE_def}
	    {\bs F}^{(t)} &= \underset{{\bs F}^\prime \in \mathbb{R}^{S\times S}}{\arg\!\min}~\mathbb{E}\big[ \| {\bs x}^{(t)} - {\bs F}^\prime{\bs q}^{(t)}\|^2\big] ={\bs R}_{{\bs x}^{(t)}{\bs q}^{(t)}} {\bs R}_{{\bs q}^{(t)}}^{-1},
	\end{align}
    ${\bs R}_{{\bs x}^{(t)}{\bs q}^{(t)}} \triangleq \mathbb{E}\big[{\bs x}^{(t)}({\bs q}^{(t)})^{\sf T}\big]$, and ${\bs R}_{{\bs q}^{(t)}} \triangleq \mathbb{E}\big[{\bs q}^{(t)}({\bs q}^{(t)})^{\sf T}\big]$.
    Note that ${\bs d}^{(t)}$ and ${\bs x}^{(t)}$ are uncorrelated by the Bussgang decomposition while ${\bs  R}_{{\bs q}^{(t)}} = \psi_Q {\bs I}_{S}$ and ${\bs  R}_{{\bs d}^{(t)}} = \big(\psi_Q - \gamma_{Q}^2\big)  {\bs I}_{S}$.
    Along with \eqref{eq:Bussgang}, the cross-correlation matrix ${\bs R}_{{\bs x}^{(t)}{\bs q}^{(t)}}$ and the correlation matrix ${\bs R}_{{\bs q}^{(t)}}$ are computed as 
	\begin{align}
    	{\bs R}_{{\bs x}^{(t)}{\bs q}^{(t)}} &= \gamma_Q  {\bs R}_{{\bs x}^{(t)}} = \gamma_Q  {\bs I}_S,~~\text{and}~~
    	{\bs R}_{{\bs q}^{(t)}} = \gamma_Q^2 {\bs R}_{{\bs x}^{(t)}} + {\bs R}_{{\bs d}^{(t)}} = \psi_Q {\bs I}_{S}, \nonumber
    \end{align}
	respectively.
 Utilizing the above result, the LMMSE estimate of ${\bs x}^{(t)}$ is determined as
	\begin{align}\label{eq:LMMSE_x}
	   \hat{\bs x}^{(t)} &= \frac{\gamma_Q}{{ \psi_Q}}{\bs q}^{(t)}.
	\end{align}
	Then, the corresponding MSE is computed as
	\begin{align}
         &\mathbb{E} \big[ \|{\bs x}^{(t)} - \hat{\bs x}^{(t)} \|^2 \big] ={\rm Tr}\big[{\bs R}_{{\bs x}^{(t)}} \big]- {\rm Tr}\big[ {\bs R}_{{\bs x}^{(t)}{\bs q}^{(t)}}  {\bs R}_{{\bs q}^{(t)}}^{-1} {\bs R}_{{\bs q}^{(t)}{\bs x}^{(t)}}\big] = S \bigg(1 -\frac{\gamma_Q^2}{ \psi_Q}\bigg).
    \end{align}	
	Next, by applying the inverse of the linear transformation in \eqref{eq:transform} to the LMMSE estimate $\hat{\bs x}^{(t)}$, an estimate of the normalized value vector ${\bs v}^{(t)}$ is given by $\hat{\bs v}^{(t)} = {\bs U}^{\sf T}\hat{\bs x}^{(t)} =  \frac{\gamma_Q}{{ \psi_Q}} {\bs U}^{\sf T}\hat{\bs q}^{(t)}$.
	Finally, by applying the inverse of the normalization in \eqref{eq:normalize} to $\hat{\bs v}^{(t)}$, an estimate of the value vector ${\bs g}_S^{(t)}$ is obtained as
    \begin{align}\label{eq:g_hat_s}
    	\hat{\bs g}_{S}^{(t)} = {\sqrt{{\nu}_{S}^{(t)}}}\hat{\bs v}^{(t)} + {\mu}_{S}^{(t)} {\bs 1}_S = \frac{\gamma_Q}{{ \psi_Q}}\sqrt{ {\nu}_S^{(t)}} {\bs U}^{\sf T}{\bs q}^{(t)} + {\mu}_{S}^{(t)}{\bs 1}_S.
	\end{align}
	Then, the corresponding MSE is computed as
	\begin{align}\label{eq:MSE_gs}
         \mathbb{E} \big[ \|{\bs g}_{S}^{(t)} - \hat{\bs g}_{S}^{(t)} \|^2 \big]
        &= \nu_{S}^{(t)}   \mathbb{E} \big[ \| {\bs U}^{\sf T}({\bs x}^{(t)}   -\hat{\bs x}^{(t)} ) \|^2 \big] = \nu_{S}^{(t)}  S \bigg(1 -\frac{\gamma_Q^2}{ \psi_Q}\bigg).
    \end{align}	
	Note that ${\mu}_{S}^{(t)}$, ${\nu}_{S}^{(t)}$, and ${\bs q}^{(t)}$ can be constructed from the value-encoded bits. 
	Following the computation of the estimated value vector $\hat{\bs g}_{S}^{(t)}$, the model update ${\bs g}^{(t)}$ can be reconstructed by inserting the values in $\hat{\bs g}_{S}^{(t)}$ into the corresponding positions.
	In {\bf Procedure~\ref{alg:Reconstruct}}, we summarize the aforementioned reconstruction strategy of FedSpar, denoted by ${\sf Reconstruct}({\bs c}^{(t)})$, which is applied to a bit sequence ${\bs c}^{(t)}$.
	
	\begin{algorithm}[!t]
		\caption{Reconstruction Strategy of FedSpar}\label{alg:Reconstruct}
		{\small
		{\begin{algorithmic}[1]
                \REQUIRE The bit sequence ${\bs c}^{(t)}$
                \ENSURE The estimate of the value vector $\hat{\bs g}^{(t)}$
		    \PROC {$ {\sf Reconstruct}\big({\bs c}^{(t)}\big)$}
		        \STATE Construct ${\mu}_{S}^{(t)}$, ${\nu}_{S}^{(t)}$, and ${\boldsymbol q}^{(t)}$ from the value-encoded bits in ${\bs c}^{(t)}$
                \STATE $\hat{\bs v}^{(t)} = \frac{\gamma_Q}{{ \psi_Q}} {\bs U}^{\sf T}\hat{\bs q}^{(t)}$
                \STATE $\hat{\bs g}_{S}^{(t)} = {\mu}_{S}^{(t)} {\bs 1}_S + {\sqrt{{\nu}_{S}^{(t)}}}\hat{\bs v}^{(t)}$
                \STATE Find $\mathcal{P}({\bs g}_{\rm sp}^{(t)}) = \{p_1,\ldots,p_S\}$ from the position-encoded bits in ${\bs c}^{(t)}$
                \STATE Initialize $\hat{\bs g}^{(t)}$ as an $N$-dimensional vector with zero entries
                \STATE Insert the $i$-th entry of $\hat{\bs g}_{S}^{(t)}$ to the $p_i$-th entry of $\hat{\bs g}^{(t)}$
                \STATE \textbf{return} $\hat{\bs g}^{(t)}$
            \ENDPROC
		\end{algorithmic}}}
	\end{algorithm}

    \subsection{Error Feedback Strategy of FedSpar}\label{Sec:Feedback}
    We also leverage an error feedback strategy to compensate for errors caused by the compression and reconstruction strategies of FedSpar.  
    Let ${\bs \Delta}_k^{(t)} \in \mathbb{R}^N$ be a local residual error defined as the difference between the local model update ${\bs g}_k^{(t)}$ and its reconstruction $\hat{\bs g}_k^{(t)}$ at device $k\in\mathcal{K}_t$, where ${\boldsymbol \Delta}_{k}^{(0)} = {\bf 0}_N$.
    Our strategy involves storing ${\bs \Delta}_k^{(t)}$ at device $k$ in iteration $t$, to retain the residual effect of the compression and reconstruction errors in the current iteration. 
    Then, the stored residual error $\bs{\Delta}_k^{(t)}$ is added to the local model update in the next participation iteration of device $k$ (i.e., Steps 10 and 13 in {\bf Procedure 3}).
    This strategy allows the devices to {\em partially} compensate for the compression and reconstruction errors in the previous participating iteration.
    The discounting factor $\kappa$ in Step 13 of {\bf Procedure~\ref{alg:FedSpar}} is used to penalize the delayed local residual error caused by the infrequent device participation.
    Note that a similar strategy was also adopted in the literature \cite{SQCS:1,SQCS:2,FedQCS,FedVQCS,D-DSGD,F. Sattler:2019} by considering only the effect of sparsification.
    In this work, we further extend the strategy in \cite{SQCS:1,SQCS:2,FedQCS,FedVQCS,D-DSGD,F. Sattler:2019} beyond the effect of sparsification to include the effects of projection, quantization, and reconstruction. 	
    The advantage of our error feedback strategy will be analyzed and evaluated in Sec.~\ref{Sec:Opt} and Sec.~\ref{Sec:Simul}, respectively.

    \subsection{Summary}\label{Sec:Summary}
    The overall procedure of FedSpar, when employing the gradient descent algorithm with a learning rate $\eta^{(t)}$ for updating the global parameter vector, is summarized in {\bf Procedure~\ref{alg:FedSpar}}. 
    Note that Step 20 can vary depending on the update rule of the employed optimizer. 
    In Steps 10 and 13, we adopt the error feedback strategy explained above.
    The determination of the sparsification level $S$ and quantization level $Q$ will be discussed in Sec.~\ref{Sec:Parameter}.

    \begin{algorithm}[t]
        \caption{The Proposed FedSpar Framework}\label{alg:FedSpar}
        {\small
        {\begin{algorithmic}[1]
            \REQUIRE The initial parameter vector ${\boldsymbol w}^{(1)}$
            \ENSURE The trained parameter vector ${\boldsymbol w}^{(T)}$
            \FOR {$t=1$ to $T$}
                \STATE \!\!{\em At the wireless devices:}
                \FORP {Each device $k\in\mathcal{K}$}
                    \IF{$k\in\mathcal{K}_t$}
                        \STATE Download ${\boldsymbol w}^{(t)}$ from the PS.
                        \STATE ${\boldsymbol g}_k^{(t)} = {\sf LocalUpdate}({\boldsymbol w}^{(t)}, \mathcal{D}_k)  + {\boldsymbol \Delta}_k^{(t-1)}$
                        \STATE Determine $(S_k,Q_k)$ such that $B_{\rm tot}\leq {C}_k N$
                        \STATE ${\boldsymbol c}_k^{(t)} = {\sf Compress}({\boldsymbol g}_k^{(t)};S_k,Q_k)$
                        \STATE $\hat{\boldsymbol g}_{k}^{(t)} = {\sf Reconstruct}({\boldsymbol c}_{k}^{(t)})$
                        \STATE ${\boldsymbol \Delta}_k^{(t)}= {\boldsymbol g}_k^{(t)} - \hat{\boldsymbol g}_{k}^{(t)}$
                        \STATE Transmit ${\boldsymbol c}_k^{(t)}$ to the PS
                    \ELSE
                        \STATE ${\boldsymbol \Delta}_k^{(t)} \leftarrow \kappa{\boldsymbol \Delta}_k^{(t-1)}$
                    \ENDIF
                \ENDFORP
                \STATE \!\!{\em At the parameter server:}
                \STATE Receive $\{{\boldsymbol c}_k^{(t)}\}_{k \in\mathcal{K}_t}$ from the wireless devices 
                \STATE $\hat{\boldsymbol g}_k^{(t)} = {\sf Reconstruct}({\boldsymbol c}_k^{(t)})$,~$\forall k \in\mathcal{K}_t$
                \STATE ${\boldsymbol g}_{\rm PS}^{(t)} = \sum_{k \in\mathcal{K}_t} \rho_k\hat{\boldsymbol{g}}_{k}^{(t)}$
                \STATE $\boldsymbol{w}^{(t+1)} \leftarrow {\boldsymbol w}^{(t)} -   \eta^{(t)} {\boldsymbol g}_{\rm PS}^{(t)}$
                \STATE Broadcast $\boldsymbol{w}^{(t+1)}$ to the wireless devices
            \ENDFOR
        \end{algorithmic}}} 
    \end{algorithm}

    {\bf Remark 1 (Complexity Reduction via Parallel Compression and Reconstruction):}
    Notably, the proposed FedSpar framework exhibits certain limitations inferred by significant computational overhead when the number of global model parameters is extremely large. 
    To overcome this limitation, FedSpar can be modified by adopting {\em parallel} compression and reconstruction, as done in \cite{FedQCS,FedVQCS}.
    This can be simply done by dividing a local model update into multiple low-dimensional sub-vectors and then by compressing and reconstructing these sub-vectors in parallel.
	Let $L$ be the number of the sub-vectors and let $N^\prime = \frac{N}{L}$ be the dimension of each sub-vector. 
        A simple yet effective strategy for constructing these $L$ sub-vectors is to randomly shuffle (permute) the entries of the local model update and then divide the entries in order.
        With this strategy, the most significant entries of the local model update can be evenly distributed across different sub-vectors. 
        Then, the $L$ sub-vectors are compressed in parallel by applying the compression strategy in Sec.~\ref{Sec:Compression} to each sub-vector. 
        The parallel compression strategy effectively reduces the computational complexity of the compression process.
        This is because the sparsification step utilizing the above strategy finds the $S^\prime$ largest entries among $N^\prime$ entries, which is repeated for $L$ times, instead of finding the $S$ largest entries among $N$ entries.
        In addition, the computational complexity of the linear transformation in \eqref{eq:transform} is also reduced from $S^2$ to $L(S^\prime)^2 = S^2/L$.
        Meanwhile, the bit overhead of the parallel compression is still similar to that of the original compression in Sec.~\ref{Sec:Compression} because $L \log_2 \binom{N^\prime}{S^\prime} \approx  L S^\prime \log_2(N^\prime/S^\prime) = S \log_2(N/S) \approx \log_2 \binom{N}{S}$ when $S/N\ll 1$.
        Thanks to the parallel compression, the $L$ sub-vectors can also be reconstructed in parallel, by applying the strategy in Sec.~\ref{Sec:Reconstruction} to reconstruct each of the sub-vectors.
        In this case, the computational complexity of the value decoding reduces from $S^2$ to $L(S^\prime)^2 = S^2/L$, as can be seen from \eqref{eq:g_hat_s}.

    \section{Convergence Analysis and Optimization of FedSpar}\label{Sec:Opt}
    In this section, we analyze the convergence rate of the proposed FedSpar framework summarized in {\bf Procedure~\ref{alg:FedSpar}}.
    On the basis of the convergence analysis, we also optimize two key parameters of FedSpar, which are (i) the sparsification level $S$ and (ii) the quantization level $Q$, for maximizing the convergence rate under a given uplink channel capacity.
    
    \subsection{Convergence Analysis}
    We analyze the convergence rate of FedSpar in {\bf Procedure~\ref{alg:FedSpar}}, which employs the gradient descent algorithm for updating the global parameter vector.  
    For this, we make the following assumptions:
    \vspace{1mm}
    
    {\em Assumption 1:}
    The loss function $F({\boldsymbol w})$ is $\beta$-smooth and is lower bounded by some constant $F(\boldsymbol{w}^{\star})$, i.e.,  $F(\boldsymbol{w})\geq F(\boldsymbol{w}^{\star})$, $\forall \boldsymbol{w}\in \mathbb{R}^{{N}}$.
    
    {\em Assumption 2:}
    The stochastic gradient is unbiased, i.e., $\mathbb{E}\big[\nabla F_k^{(t,e)}({\boldsymbol w})\big] = \nabla F_k({\boldsymbol w})$.
    Also, the expected squared norm of the stochastic gradient is uniformly bounded, i.e.,
    $\mathbb{E}\big[\big\|\nabla F_k^{(t,e)}({\boldsymbol w})\big\|^2\big] \leq \epsilon^2$, $\forall t,k,e$.

     Assumptions 1 and 2 have been widely adopted in the literature for mathematical tractability of the convergence analysis with the consideration of {\em non-IID} data distribution as in \cite{SignSGD, VeQ:1,Convergence,FedCOM}.
    \begin{lem}\label{lemma:Error_norm}
    Under Assumption 2 with $E=1$ and $\rho_k=\frac{1}{M}$, the expected squared norms of the residual errors at device $k \in\mathcal{K}_t$ are bounded:
    \begin{align}        \mathbb{E}\big[\big\|\boldsymbol{\Delta}_k^{(t)}\big\|^2\big] &\leq  \delta_k^2 \triangleq \frac{4(1-{\zeta}_k)\epsilon^2}{{\zeta}_k^2},  \label{eq:Error_K_norm}
    \end{align}
    where ${\zeta}_k =  \frac{\gamma_{Q_k}^2S_k}{ \psi_{Q_k}N}$, provided that the sparsification level of device $k$ is $S_k$, and the quantization level of device $k$ is $Q_{k}$.
    \end{lem}
    
    \begin{IEEEproof}
        See Appendix~\ref{Apdx:Lemma1}.
    \end{IEEEproof}
    \vspace{1mm}

    Under Assumptions 1--2 and Lemma 1, we characterize the convergence rate of FedSpar as given in the following theorem:
    \vspace{1mm}
    
    \begin{thm}\label{Thm1}
        Under Assumptions 1--2 and Lemma 1, FedSpar in {\bf Procedure~\ref{alg:FedSpar}} with $E=1$, $\rho_k=\frac{1}{M}$, $\kappa=1$, and $\eta^{(t)} = \frac{2}{\sqrt{T}}$ satisfies the following bound:
        \begin{align}\label{eq:Thm2}
            &\mathbb{E} \left[\frac{1}{T} \sum_{t=1}^T  \| \nabla F(\boldsymbol{w}^{(t)}) \|^2\right] \leq
        \frac{1}{\sqrt{T}}\left\{F({\boldsymbol w}^{(1)}) - F({\boldsymbol w}^{\star}) + 2\beta \epsilon^2 
        + \frac{4K^2}{M^2 \sqrt{T}}\beta^2\max_k\delta_k^2\right\}.
        \end{align}
    \end{thm}
    
    \begin{IEEEproof}
        See Appendix~\ref{Apdx:Thm1}.
    \end{IEEEproof}
    \vspace{1mm}
    
    Theorem 1 demonstrates that under Assumptions 1--2 and Lemma 1, FedSpar converges to a stationary point of a non-convex loss function. 
    This theorem also shows that the convergence rate of FedSpar has an order $\mathcal{O}({1}/{\sqrt{T}})$ which is the same as the order achieved by a gradient-descent-based algorithm for a non-convex loss function \cite{SignSGD, FedVQCS, FedQCS}. 
    This implies that our analytical result also coincides with the existing results in the FL literature. 
    The convergence rate of FedSpar improves as $\delta_k^2$ decreases, which corresponds to the upper bound of $\mathbb{E}[\|{\boldsymbol \Delta}_k^{(t)} \|^2]$. 
    As can be seen in Lemma~1, the residual error ${\boldsymbol \Delta}_k^{(t)}$ vanishes as $C_k$ increases.
    Therefore, Theorem~1 implies that the convergence rate improves as the uplink capacity increases, as naturally expected. 

    \subsection{Parameter Optimization}\label{Sec:Parameter}
    The compression strategy of FedSpar has two design parameters: (i) the sparsification level $S$ and (ii) the quantization level $Q$. 
    These parameters affect not only the bit overhead required for the transmission (see Section III-A), but also the convergence rate of FedSpar (see Theorem~1). 
    In particular, Theorem~1 shows that the convergence rate can be maximized if the expected squared norm of the residual error, denoted by $\mathbb{E}[\|{\boldsymbol \Delta}_k^{(t)} \|^2]$, is minimized.
    Motivated by this, we optimize both $Q$ and $S$ for minimizing the expected squared norm of the residual error under the bit-overhead constraint of $B_{\rm tot} \leq C_kN$ at device $k \in \mathcal{K}_t$.   
    %
    Let $\hat{\bs g}^{(t)}$ be the reconstruction of the local model update ${\bs g}^{(t)}$ at device $k \in \mathcal{K}_t$ determined by the reconstruction strategy of FedSpar.
    Then, an optimization problem to find the optimal sparsification and quantization levels, which minimize the expected squared norm of the residual error ${\bs \Delta}^{(t)} = {\bs g}^{(t)} - \hat{\bs g}^{(t)}$, is formulated as
    \begin{align}\label{eq:problem}
        (S^\star, Q^\star)= \underset{S \leq N, Q \in \mathcal{Q}}{\arg\!\min}~  \mathbb{E} \big[ \|{\bs g}^{(t)} - \hat{\bs g}^{(t)} \|^2\big] ~~\text{s.t.}~~  B_{\rm tot} \leq CN,
    \end{align}
    where $\mathcal{Q} \triangleq \{2,\ldots,Q_{\rm max}\}$, and $Q_{\rm max}$ is the maximum quantization level considered in the system.
    In particular, when computing the MSE in \eqref{eq:problem}, we model the transformed value vector ${\bs x}^{(t)}$ in \eqref{eq:transform} as a random vector distributed as ${\bs x}^{(t)} \sim \mathcal{N}({\bs 0}_S, {\bs I}_S)$, which has been already justified in Sec.~\ref{Sec:Compression} for large $S$. 
    Moreover, the value vector ${\bs g}_S^{(t)}$ is expressed as 
    \begin{align}
        {\bs g}_S^{(t)} = \sqrt{ {\nu}_S^{(t)}} {\bs U}^{\sf T}{\bs x}^{(t)} + {\mu}_{S}^{(t)}{\bs 1}_S,
    \end{align}
    from \eqref{eq:normalize} and \eqref{eq:transform}. 
    Consequently, we can model ${\bs g}_S^{(t)}$ as a random vector, while treating  ${\nu}_S^{(t)}$ and ${\mu}_{S}^{(t)}$ as known constants\footnote{Both ${\nu}_S^{(t)}$ and ${\mu}_{S}^{(t)}$ are deterministic functions of $S$ and ${\bs g}^{(t)}$ by the definitions in \eqref{eq:mean} and \eqref{eq:variance}.}. 
    Except the entries in ${\bs g}_S^{(t)}$, the remaining entries of ${\bs g}^{(t)}$ are treated as known constants.  
    
    In FedSpar, an estimation error only occurs in the positions of the $S$ largest entries of ${\bs g}^{(t)}$, while a sparsification error occurs in the remaining positions. 
    Considering this fact, the MSE $\mathbb{E}\big[ \|{\bs g}^{(t)} - \hat{\bs g}^{(t)} \|^2 \big]$ for given $S$ and $Q$ can be rewritten as
    \begin{align}\label{eq:MSE}
         \mathbb{E} \big[ \|{\bs g}^{(t)} - \hat{\bs g}^{(t)} \|^2 \big] 
        &= \sum_{i=S+1}^{N} \big|g_{{\rm mag},i}^{(t)} \big|^2 +  \mathbb{E} \big[ \|{\bs g}_S^{(t)} - \hat{\bs g}_{S}^{(t)} \|^2 \big] \nonumber \\
        &\overset{(a)}{=}\sum_{i=S+1}^{N} \big|g_{{\rm mag},i}^{(t)} \big|^2
        + S \nu_{S}^{(t)}\bigg(1- \frac{\gamma_Q^2}{ \psi_Q} \bigg) \nonumber\\
        &\overset{(b)}{=} \|{\bs g}^{(t)}\|^2 - \frac{\gamma_Q^2}{ \psi_Q}  \|{\bs g}_S^{(t)}\|^2 - S\big({\mu}_{S}^{(t)} \big)^2 \bigg(1- \frac{\gamma_Q^2}{ \psi_Q} \bigg),
    \end{align} 
    where $g_{{\rm mag},i}^{(t)}$ is the $i$-th largest entry of ${\bs g}^{(t)}$ in terms of magnitude, $(a)$ follows from the MSE analysis in \eqref{eq:MSE_gs}, and $(b)$ follows from the definition in \eqref{eq:variance}.  
    The term $\|{\bs g}^{(t)}\|^2$ in \eqref{eq:MSE} does not depend on the choice of $S$ and $Q$. Thus, from \eqref{eq:MSE}, the problem in \eqref{eq:problem} can be rewritten as 
    \begin{align}\label{eq:problem2}
        (S^\star, Q^\star)= \underset{S \leq N, Q \in \mathcal{Q}}{\arg\!\max}~   \frac{\gamma_Q^2}{ \psi_Q}  \|{\bs g}_S^{(t)}\|^2 + S\big({\mu}_{S}^{(t)} \big)^2 \bigg(1- \frac{\gamma_Q^2}{ \psi_Q} \bigg) ~~\text{s.t.}~~  B_{\rm tot} \leq CN.
    \end{align}  
    The bit-overhead analysis in \eqref{eq:total_bit} implies that if $C_k\leq 1$, the sparsification level $S$ should be less than $\frac{N}{2}$ to satisfy $B_{\rm tot} \leq C_kN$.
    Moreover, if $S \leq \frac{N}{2}$, the bit overhead $B_{\rm tot}$ is an increasing function of $S$ for a fixed quantization level $Q$.
    These facts imply that the maximum sparsification level $S_Q^{\rm max}$ for a given $Q \in \mathcal{Q}$ can be uniquely determined by solving the following problem:
    \begin{align}\label{eq:S_max}
        S_{Q}^{\rm max} = \underset{S \leq \frac{N}{2}}{\arg\!\max}~S~~
        \text{s.t.}~~S\log_2Q + 64 + \log_2 \binom{N}{S} \leq C_kN,
    \end{align} 
    which can be readily solved by using numerical methods. 
    Then, the constraint of $B_{\rm tot}\leq C_kN$ is substituted by $S \leq S_Q^{\rm max}$ for the given quantization level $Q \in \mathcal{Q}$. 
    Therefore, the optimal sparsification level $S_Q^\star$ for a given $Q \in \mathcal{Q}$ can be determined by solving the following problem:
    \begin{align}\label{eq:S_problem}
        S_Q^\star = \underset{S \leq S_Q^{\rm max}}{\arg\!\max}~ \frac{\gamma_Q^2}{ \psi_Q}  \|{\bs g}_S^{(t)}\|^2 + S\big({\mu}_{S}^{(t)} \big)^2 \bigg(1- \frac{\gamma_Q^2}{ \psi_Q} \bigg).
    \end{align}    
    Similarly, the optimal quantization level $Q^\star$ can be determined by solving the following problem:
    \begin{align}\label{eq:Q_problem}
        Q^\star = \underset{Q\in \mathcal{Q}}{\arg\!\max}~  \left\{ \max_{S\leq S_Q^{\rm max}}~  \frac{\gamma_Q^2}{ \psi_Q}  \|{\bs g}_S^{(t)}\|^2 + S\big({\mu}_{S}^{(t)} \big)^2 \bigg(1- \frac{\gamma_Q^2}{ \psi_Q} \bigg) \right\}.
    \end{align}
    Note that the objective function in \eqref{eq:Q_problem} is a deterministic function of ${\bs g}^{(t)}$ for given $Q$ and $S$.
    Therefore, the above problem can be solved by comparing the objective values of \eqref{eq:Q_problem} computed for all $S\leq S_Q^{\rm max}$ and $Q\in\mathcal{Q}$ based on the information of ${\bs g}^{(t)}$.

    The aforementioned exhaustive approach, however, may incur significant computational overhead when both $|\mathcal{Q}|$ and $S_Q^{\rm max}$ are large. 
    To circumvent this challenge, we reformulate the problem in \eqref{eq:Q_problem} by assuming that the mean $\mu_S^{(t)}$ is zero which is likely to be true for large $S$ by the law of large numbers.   
    Under this assumption, the optimal sparsification level $S_Q^\star$ for the given quantization level $Q \in \mathcal{Q}$ is determined as
    \begin{align}\label{eq:S_Q3}
        S_Q^\star \approx \underset{S \leq S_Q^{\rm max}}{\arg\!\max}~  
        \frac{\gamma_Q^2}{ \psi_Q} \|{\bs g}_{S}^{(t)}\|^2
        \overset{(a)}{=} S_{Q}^{\rm max},
    \end{align}
    where $(a)$ holds because $\|{\bs g}_{S}^{(t)}\|^2$ is an increasing function of $S$. 
    Consequently, the optimal quantization level $Q^\star$ can be determined by solving the following problem:
    \begin{align}\label{eq:Q_opt}
        Q^\star 
        \approx \underset{Q\in \mathcal{Q}}{\arg\!\max}~  \frac{\gamma_Q^2}{ \psi_Q}\|{\bs g}_{S_{Q}^{\rm max}}^{(t)}\|^2
        = \underset{Q\in \mathcal{Q}}{\arg\!\max}~ \frac{\gamma_Q^2}{ \psi_Q} \sum_{i=1}^{S_Q^{\rm max}} \big|g_{{\rm mag},i}^{(t)}\big|^2.
    \end{align}
    As can be seen in \eqref{eq:Q_opt}, determining $Q^\star$ only requires the computation of the objective values in \eqref{eq:Q_opt} for all  $Q\in\mathcal{Q}$. This leads to significantly lower computational overhead for solving the problem in \eqref{eq:Q_opt} compared to directly solving the problem in \eqref{eq:Q_problem}.
    Once the optimal quantization level $Q^\star$ is determined, the optimal sparsification level is directly determined as $S^\star = S_{Q^\star}^{\rm max}$ from \eqref{eq:S_Q3} without additional computation. 
    Our parameter optimization strategy can be leveraged in Step 7 of {\bf Procedure~\ref{alg:FedSpar}} for uplink transmission at device $k \in \mathcal{K}_t$.

    \section{Simulation Results}\label{Sec:Simul}

    
   In this simulation, we consider FL for an image classification task using the publicly accessible MNIST \cite{MNIST} and CIFAR-$10$ \cite{CIFAR-10} datasets. Details of the learning scenarios for each dataset are described below.
    \begin{itemize}
        \item {\em MNIST:}
        A fully-connected neural network is employed as the global model on the PS. 
        This network consists of $784$ input nodes, single hidden layer with $20$ hidden nodes, and $10$ output nodes. 
        The activation functions of the hidden and output layers are set as the rectified linear unit (ReLU) and softmax functions, respectively. 
        For updating the global parameter vector (i.e., Step 20 in {\bf Procedure~\ref{alg:FedSpar}}), the ADAM optimizer in \cite{Adam} is employed with an initial learning rate $0.01$ and the cross-entropy loss function.
        To consider a {\em non-IID} distribution in the MNIST dataset, the local training dataset for device $k$, namely $\mathcal{D}_k$, is determined by randomly selecting $1000$ training data samples from one class.
        The mini-batch size in each iteration is $|\mathcal{D}_k^{(t,e)}| = 10$, and the number of local iterations is $E=1$. 
        The system consists of a total of $K=50$ devices, out of which $M=20$ devices participate in each iteration, and the number of the iterations is $T=100$.

        
        \item {\em CIFAR-10:}
        The ShuffleNet in \cite{ShuffleNet} is employed as the global model on the PS.
        For updating the global parameter vector (i.e., Step 20 in {\bf Procedure~\ref{alg:FedSpar}}), the ADAM optimizer in \cite{Adam} is employed with an initial learning rate 0.001 and the cross-entropy loss function.
        To consider a {\em non-IID} distribution in the CIFAR-$10$ dataset, the local training dataset for device $k$, namely $\mathcal{D}_k$,  is determined by a Dirichlet distribution with concentration parameter $\alpha = 0.6$.
        For updating the local parameter vector in \eqref{eq:local_model_update}, the mini-batch gradient descent algorithm is employed with $\gamma^{(t)} = 0.01$ for the cross-entropy loss function.
        The mini-batch size in each iteration is $|\mathcal{D}_k^{(t,e)}| = 50$, and the number of local iterations is $E=10$. 
        The system consists of a total of $K=50$ devices, out of which $M=20$ devices participate in each iteration, and the number of the iterations is $T=100$.
        
    \end{itemize}
    For performance comparisons, we consider the following FL frameworks:

	\begin{itemize}
	    \item {\em Vanilla FL:}
	    This is the vanilla FL framework that assumes perfect uplink transmission without considering the effect of a limited uplink channel capacity.
	    
	    \item {\em FedSpar:}
	    This is the proposed FL framework summarized in {\bf Procedure~\ref{alg:FedSpar}} with the parameter optimization in Sec.~\ref{Sec:Parameter}. 
            We set $Q_{\rm max}=16$ while setting $L=1$ for the MNIST dataset and $L=603$ for the CIFAR-$100$ dataset as these values provide a reasonable trade-off between the classification accuracy and computational complexity of FedSpar.
	    
	    \item {\em FedVQCS:}
	    FedVQCS is the communication-efficient FL framework developed in \cite{FedVQCS}. All       simulation parameters are set as described in Sec. V in \cite{FedVQCS}.                    We set $(G,B)=(3,10)$ for the MNIST dataset and $(G,B) = (3,9302)$ for the CIFAR-$100$ dataset.
	    
	    \item {\em FedQCS:}
	    FedQCS is the communication-efficient FL framework developed in \cite{FedQCS} when adopting the aggregate-and-estimate strategy in the reconstruction process. 
    We set $Q_k = 1$, $R_k = 1/C_k$, and $S_k = \max \left\{ S \geq 1 \big| R_k < \frac{GN}{{2 BK     S \log(GN/(BK S))}} \right\}$ for $k\in\mathcal{K}$, where $(G,B)=(3,10)$ for the          MNIST dataset and $(G,B) = (3,9302)$ for the CIFAR-$100$ dataset. 
     
	    
	    \item {\em HighVQ:}
	    HighVQ is the communication-efficient FL framework developed in \cite{VeQ:3}. We set $Q_k = C_k$ and also set the dimension of each partition for vector quantization as the largest integer $L$ such that $L2^{C_kL}\leq2^{15}$ for device $k\in\mathcal{K}$.
	    
	    \item {\em D-DSGD:} 
	    D-DSGD is the communication-efficient FL framework developed in \cite{D-DSGD}. We set the number of the non-zero entries for device $k\in\mathcal{K}$ as $S_k$ such that $C_k{N} = \log_2\big(_{S_k}^{N}\big) + 33$.
     
     
    \end{itemize}
    We assume that the parameters of batch normalization for all FL frameworks are perfectly transmitted without requiring communication overhead.
        \subsection{Homogeneous Communication Scenario}\label{Sec:Homogeneous} 
        In this scenario, we assume that all devices have equal capacities by setting $C_k = C$ for all $k\in\mathcal{K}$, where $C_k$ has been defined as the number of bits per local model entry allowed for transmitting the local model update at device $k$. 
        

    \begin{table}[t]
        \renewcommand{\arraystretch}{0.7}
        \centering
        \caption{Classification accuracy vs. uplink communication overhead for different FL frameworks for the MNIST and CIFAR-$10$ datasets.} \label{table:Compare}
        \begin{tabular}{|c | c | c | c| c|| c | c | c| c|}
            \hline
            Dataset & \multicolumn{4}{c||}{MNIST} & \multicolumn{4}{c|}{CIFAR-$10$} \\ \hline 
            Compression Ratio & 320 $\times$ & 160 $\times$ & 80 $\times$ & 1 $\times$ & 320 $\times$ & 160 $\times$ & 80 $\times$ & 1 $\times$\\ \hline
            $C$ (bits/entry) & 0.1 & 0.2 & 0.4 & 32 & 0.1 & 0.2 & 0.4 & 32 \\ \hline
            Bit overhead & 0.20 KB & 0.40 KB  & 0.80 KB & 63.64 KB & 10.84 KB & 21.68 KB & 43.36 KB & 3.47 MB \\ \hline \hline
             Vanilla  FL & - & - & - & {\bf 90.67}\% & - & - & - & {\bf 64.49}\% \\ \hline
            \multirow{2}{*}{FedSpar} & {\bf 86.53}\% & {\bf 88.66}\% & {\bf 89.70}\% & \multirow{2}{*}{-} & {\bf 59.28}\% & {\bf 62.12}\% & {\bf 63.15}\% & \multirow{2}{*}{-}\\ 
            & (0.9\%$^\star$)\tnote{1} & (2.0\%$^\star$) & (4.5\%$^\star$) &  & (1.0\%$^\star$) & (2.2\%$^\star$) & (5.0\%$^\star$) &\\ \hline
            FedSpar & 80.44\% & 84.46\% & 87.46\% & \multirow{2}{*}{-} & 38.26\% & 47.39\% & 48.95\% & \multirow{2}{*}{-}\\
            (No error feedback) & (0.8\%$^\star$) & (1.7\%$^\star$) & (4.0\%$^\star$) & & (0.9\%$^\star$) &(2.0\%$^\star$) &(4.5\%$^\star$) &\\ \hline            
            \multirow{2}{*}{FedVQCS} & 85.48\% & 86.46\% & 87.84\% & \multirow{2}{*}{-} & 53.01\% & 56.37\% & 58.64\% & \multirow{2}{*}{-}\\
            & (5.9\%$^\star$) & (6.2\%$^\star$) & (6.6\%$^\star$) & & (5.5\%$^\star$) & (6.7\%$^\star$) & (7.2\%$^\star$) &\\ \hline 
            \multirow{2}{*}{FedQCS} & 77.09\% & 80.45\% & 84.92\%  & \multirow{2}{*}{-} & 48.67\% & 53.95\% & 57.79\% & \multirow{2}{*}{-}\\
            & (0.3\%) & (0.9\%) & (2.6\%) & & (0.3\%) & (0.9\%) & (2.6\%) &\\ \hline  
            HighVQ & 79.24\% & 82.94\% & 84.25\% & - & 54.32\% & 57.81\% & 60.13\% &-\\ \hline
            \multirow{2}{*}{D-DSGD} & 80.30\% & 83.76\% & 87.05\%  & \multirow{2}{*}{-} & 59.03\% & 61.00\% & 61.20\% & \multirow{2}{*}{-}\\
            & (1.3\%) & (3.1\%) & (7.9\%) & & (1.3\%) & (3.1\%) & (7.9\%) &\\ \hline 
        \end{tabular}
    \end{table}
    
    In Table~\ref{table:Compare}, we evaluate the performance gain of the proposed FedSpar framework compared to the existing FL frameworks under the homogeneous scenario.
    To this end, we compare the classification accuracy of FedSpar with those of the existing communication-efficient FL frameworks for the MNIST and CIFAR-$10$ datasets.
    In this table, we also specify the compression ratio, bit overhead, and sparsification ratio of the uplink transmission in different FL frameworks, where the value in $(\cdot)$ represents a fixed sparsification ratio and the value in $(\cdot^\star)$ represents an {\em average} sparsification ratio. 
    Table~\ref{table:Compare} shows that for the same compression ratio, FedSpar achieves a higher classification accuracy than all the existing FL frameworks.
    In particular, for the MNIST dataset, FedSpar with $ C =0.4 $ yields only a 0.97\% decrease in classification accuracy compared to Vanilla FL.
    This result demonstrates that FedSpar reduces the uplink communication overhead in FL without causing a significant loss in the FL performance. 
    The performance gap between FedSpar with (i.e., $\kappa = 1$) and without the error feedback strategy (i.e., $\kappa = 0$) is shown to be considerable for all cases, while this gap increases as the uplink communication overhead decreases.
    This result shows the importance and efficacy of the error feedback strategy adopted in FedSpar, particularly when the uplink capacity is very limited.

    \begin{figure}  
    \centering
        {\epsfig{file=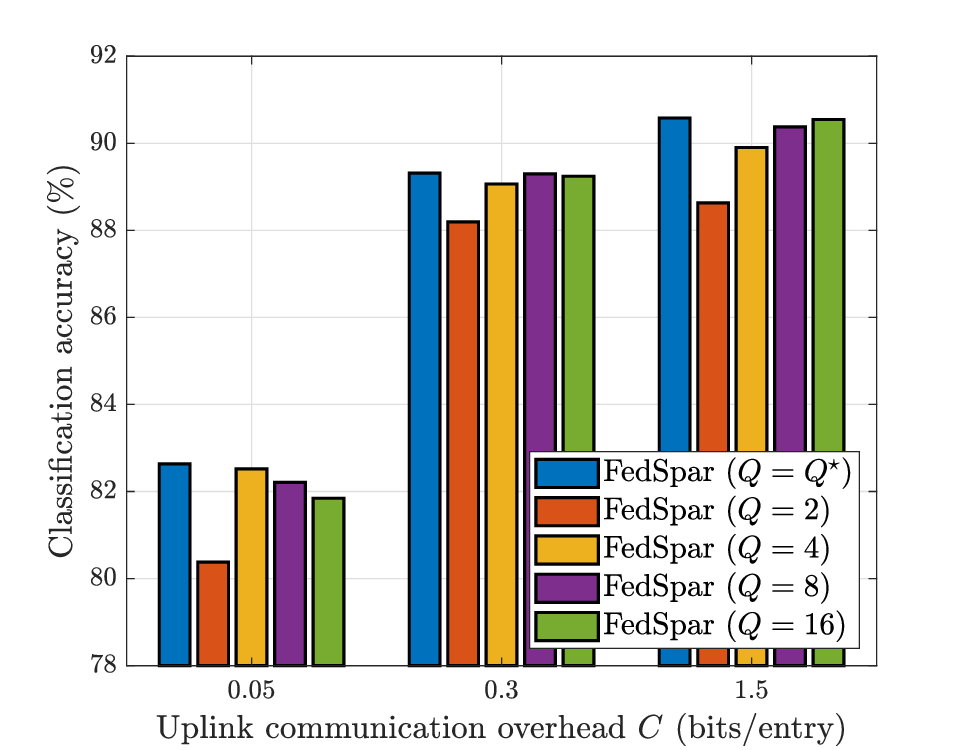, width=7.7cm}}
		\caption{Classification accuracy vs. uplink communication overhead achieved by the proposed FedSpar framework with and without the parameter optimization for the MNIST dataset when $C = \{0.05, 0.3, 1.5\}$.} \vspace{-5mm} 
		\label{fig:opt}
    \end{figure}
    
   

	In Fig.~\ref{fig:opt}, we evaluate the effectiveness of the parameter optimization in Sec.~\ref{Sec:Parameter}.
    To this end, we compare the classification accuracy of FedSpar with and without the parameter optimization for the MNIST dataset when $C = \{0.05, 0.3, 1.5\}$. 
	For FedSpar without the parameter optimization, we fix the quantization level $Q$ as one of $\{2,4,8,16\}$. 
    Fig.~\ref{fig:opt} shows that FedSpar with the parameter optimization provides the highest accuracy regardless of the uplink communication overhead.
    Meanwhile, among the fixed quantization levels, no single level provides the highest classification accuracy for all cases. 
    These results demonstrate the importance and effectiveness of the parameter optimization in Sec.~\ref{Sec:Parameter} for improving the performance of FedSpar. 
    It is also shown that the value of the best quantization level among the fixed levels decreases as the uplink communication overhead reduces. 
    This result implies that as the capacity decreases, it is more important to increase the number of transmitted non-zero entries, than to reduce the quantization error of each entry. 
    %
    %
    \begin{figure}
        \centering
        \begin{minipage}{0.48\columnwidth}
            \centering
            \vspace{4.5mm}
            {\setlength{\fboxrule}{0pt}
            \hspace{-4.5mm}
            \fbox{{\epsfig{file=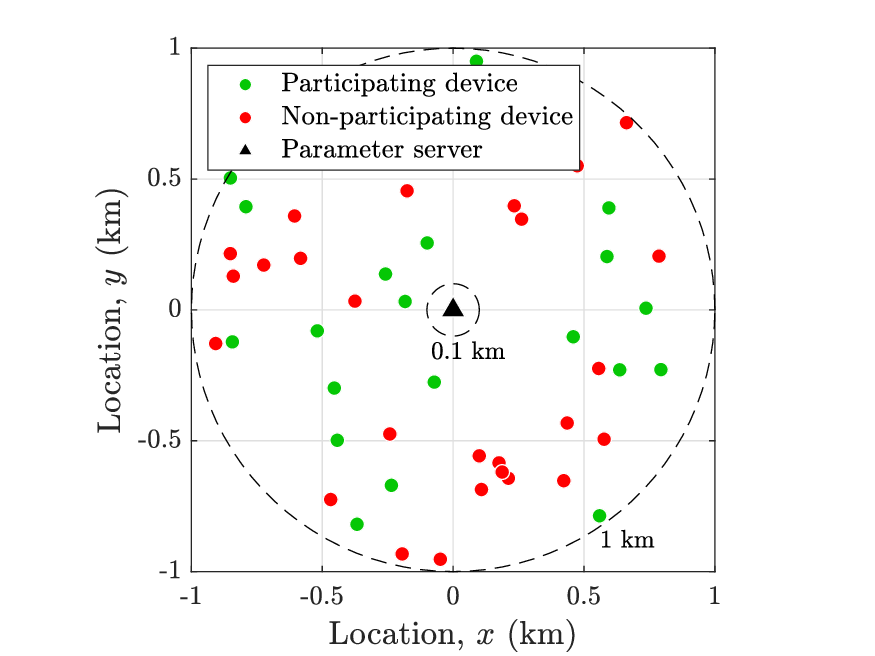, width=8cm}}}
            \caption{Illustration of the considered wireless network with $K=50$ devices.}
            \label{fig:cell}}
        \end{minipage}
         \hspace{4mm}
        \begin{minipage}{0.48\columnwidth}
            \centering
            {\setlength{\fboxrule}{0pt}
            \fbox{{\epsfig{file=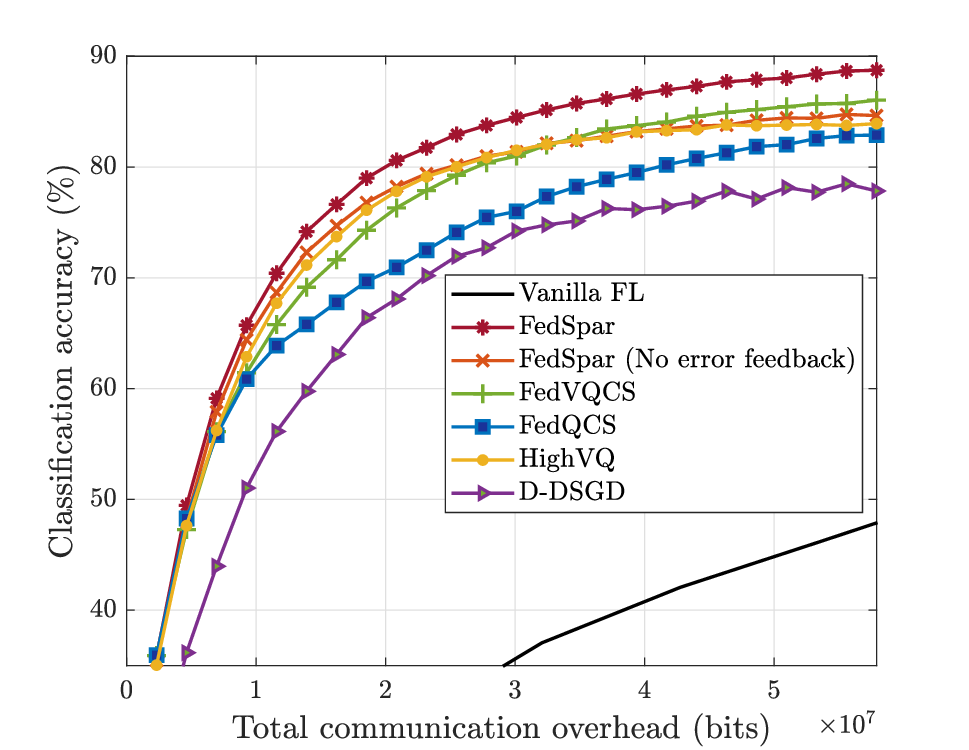, width=8cm}}}
            \caption{Classification accuracy vs. total communication overhead for different FL frameworks with the MNIST dataset.}
            \vspace{-3mm}
            \label{fig:MNIST}}
        \end{minipage}
    \end{figure}
    \subsection{Heterogeneous Communication Scenario}\label{Sec:Heterogeneous}
    In this scenario, we consider a more realistic wireless FL environment by incorporating various uplink capacities among the devices in a wireless network.
    Fig.~\ref{fig:cell} illustrates a snapshot of the considered wireless network with $K=50$ devices, where the locations of the devices are uniformly distributed between $100$ m and $1,000$ m from the center of a cell.
    To model the capacities, we employ the log-distance path loss model in \cite{PathLoss}.  Under this model, the path loss of device $k$ is determined as ${\rm PL}_{k} = A + 10\gamma_{\rm PL}\log_{10}(d_k/d_0) + Z$, where $A=20\log_{10} (4\pi d_0f_{c}/c_{{\rm light}})$ is the intercept value, $c_{{\rm light}}$ is the speed of light, $f_{c} = 2.4$ GHz is the carrier frequency, $\gamma_{\rm PL} = 4$ is the path loss exponent, $d_0 = 100$ m is the reference distance, $d_k\geq d_0$ is the distance between the PS and the device, and $Z$ is the shadow fading term that follows the Gaussian distribution with zero mean and $8.7$-dB variance. The signal-to-noise ratio (SNR) of device $k$ is defined as ${\rm SNR}_k = P_{\rm S} - {\rm PL}_{k}$ (dB), where $P_{\rm S}$ (dB) represents the scaling factor, which is set to satisfy $\mathbb{E}_k[{\rm SNR}_k] = 10$ dB. Based on the SNR ${\rm SNR}_k$, we quantify the uplink transmission bits of device $k$ as $C_kN = T_{\rm up}W\log_2(1 + {\rm SNR}_{{\rm lin},k})$ (bits),
    where $W = 1$ MHz is the bandwidth, $T_{\rm up} = 1$ ms is the uplink transmission time, and ${\rm SNR}_{{\rm lin},k} = 10^{\frac{{\rm SNR}_k}{10}}$. 

    In Fig.~\ref{fig:MNIST}, we evaluate the performance gain of the proposed FedSpar framework compared to the existing FL frameworks under the heterogeneous scenario by comparing the classification accuracy for the MNIST dataset. Our evaluation considers the total communication overhead of FL, which includes both the communication overhead for transmitting local model updates at the devices and the communication overhead for transmitting the global model at the PS. The average of $C_k$ is determined as $0.22$ bits/entry, considering an average signal-to-noise ratio of $\mathbb{E}_k[{\rm SNR}_k] = 10$ dB. Fig.~\ref{fig:MNIST} shows that FedSpar outperforms all baselines in terms of classification accuracy, while maintaining a relatively low communication overhead. This highlights the ability of the proposed framework to achieve high classification accuracy while effectively managing communication constraints in the heterogeneous wireless environment.

    \section{Conclusion}
    In this paper, we have presented a novel communication-efficient FL framework for improving the convergence rate of FL under a limited uplink capacity. The key idea behind the presented framework is to encode and transmit the values and positions of the $S$ largest entries of a local model update for uplink transmission. We have shown that our framework effectively reduces the uplink communication overhead in FL by leveraging the sparse property of the local model update. We have also minimized the compression and reconstruction errors by optimizing the quantizer design for the value encoding and developing an LMMSE-based reconstruction method. Based on the convergence analysis, we have optimized the key parameters of the presented framework for maximizing its convergence rate. Using simulations, we have demonstrated that our framework outperforms state-of-the-art FL frameworks under the same communication overhead. An important direction for future research is to modify the presented framework to support dynamic device scheduling and resource allocation which can further improve the performance of FL. Another promising research direction is to extend the presented framework by considering the heterogeneity of the local data distributions and computing capabilities of the devices. It would also be possible to improve the communication efficiency of the presented framework by using temporal correlation of model updates, as suggested in \cite{TCS}. 
    \appendices
        \section{Proof of Lemma~\ref{lemma:Error_norm}\label{Apdx:Lemma1}}
       We start by proving \eqref{eq:Error_K_norm} in Lemma 1, where the main steps of this proof are similar to the proof given in \cite{ErrorAccum}. 
       Suppose that the sparsification and quantization levels of device $k\in\mathcal{K}_t$ are set as $S_k$ and $Q_k$,  respectively. 
       From \eqref{eq:MSE} and Step 13 of {\bf Procedure 3}, the expected squared norm of the residual error at device $k \in \mathcal{K}_t$ is upper bounded as  
        \begin{align}\label{eq:errorbound1}
            \mathbb{E}\big[\big\|\boldsymbol{\Delta}_k^{(t)}\big\|^2\big] &= \mathbb{E} \big[ \|{\boldsymbol g}_k^{(t)} - \hat{\boldsymbol g}_k^{(t)} \|^2 \big]  = \mathbb{E}\left[\|{\boldsymbol g}_k^{(t)}\|^2 - \frac{\gamma_{Q_k}^2}{ \psi_{Q_k}}  \|{\boldsymbol g}_{k,S_k}^{(t)}\|^2 - S_k\big({\mu}_{k,S_k}^{(t)} \big)^2 \bigg(1- \frac{\gamma_{Q_k}^2}{ \psi_{Q_k}} \bigg) \right] \nonumber \\
            &\leq \mathbb{E}\big[ \|{\boldsymbol g}_k^{(t)}\|^2\big] - \frac{\gamma_{Q_k}^2}{ \psi_{Q_k}} \mathbb{E}\big[ \|{\boldsymbol g}_{k,S_k}^{(t)}\|^2\big], 
        \end{align}
        where ${\boldsymbol g}_{k,S_k}^{(t)}$ is the sparsified model update of device $k$ and ${\mu}_{k,S_k}^{(t)} = {\bf 1}_{S_k}^{\sf T}{\bf g}_{k,S_k}^{(t)}/S$. Since ${\boldsymbol g}_{k,S_k}^{(t)}$ consists of the $S_k$ largest entries of ${\boldsymbol g}_k^{(t)}$, we have $\|{\boldsymbol g}_{k,S_k}^{(t)}\|^2 \geq \frac{S_k}{N}\|{\boldsymbol g}_k^{(t)}\|^2$. 
        Applying this inequality into \eqref{eq:errorbound1} yields 
        \begin{align}\label{eq:errorbound2}     
            \mathbb{E}\big[\big\|\boldsymbol{\Delta}_k^{(t)}\big\|^2\big] 
            &\leq \mathbb{E}\big[ \|{\boldsymbol g}_k^{(t)}\|^2\big] - \frac{\gamma_{Q_k}^2 S_k}{ \psi_{Q_k} N} \mathbb{E}\big[ \|{\boldsymbol g}_k^{(t)}\|^2\big] = (1-\zeta_k)\mathbb{E}\big[ \|{\boldsymbol g}_k^{(t)}\|^2\big],
        \end{align}
        where ${\zeta}_k = \frac{\gamma_{Q_k}^2 S_k}{ \psi_{Q_k} N}$. 
        Note that $0< {\zeta}_k < 1$ because $\psi_{Q_k} - \gamma_{Q_k}^2>0$ and $S_k<N$ by the definitions. From Step 6 of {\bf Procedure 3}, we have
        \begin{align}\label{eq:g_bound}     
            \mathbb{E}\big[ \|{\boldsymbol g}_k^{(t)}\|^2\big] 
            &= \mathbb{E}\big[ \|{\sf LocalUpdate}(\boldsymbol{w}^{(t)},\mathcal{D}_k) + {\boldsymbol \Delta}_k^{(t-1)} \|^2\big]  \nonumber  \\
            &\overset{(a)}{\leq} (1 +\kappa)\mathbb{E}\big[\|{\boldsymbol \Delta}_k^{(t-1)}\|^2\big] + \left(1 +\frac{1}{\kappa}\right)\mathbb{E}\big[\|{\sf LocalUpdate}({\boldsymbol w}^{(t)}, \mathcal{D}_k) \|^2 \big] \nonumber \\
            &\overset{(b)}{\leq}  (1 +\kappa)\mathbb{E}\big[\|{\boldsymbol \Delta}_k^{(t-1)}\|^2\big]  + \left(1 +\frac{1}{\kappa}\right)\epsilon^2,
        \end{align}
        where $(a)$ holds for any $\kappa > 0$ by Young's inequality, and $(b)$ follows from  Assumption 2 and \eqref{eq:local_model_update} with $E=1$.
        Applying the inequality in \eqref{eq:g_bound} and $\kappa = \frac{\zeta_k}{2(1-\zeta_k)} > 0$ into \eqref{eq:errorbound2} yields
        \begin{align}\label{eq:errorbound3}   
            \mathbb{E}\big[\big\|\boldsymbol{\Delta}_k^{(t)}\big\|^2\big]
            &\leq \left(1-\frac{\zeta_k}{2}\right)\mathbb{E}\big[\|{\boldsymbol \Delta}_k^{(t-1)}\|^2\big] + (1-{\zeta}_k)\left(\frac{2}{\zeta_k} -1\right)\epsilon^2 \nonumber \\
            &\overset{(a)}{\leq} (1-{\zeta}_k)\left(\frac{2}{\zeta_k} -1\right)\epsilon^2 \sum_{i=0}^{t-1} \left(1-\frac{\zeta_k}{2}\right)^i \nonumber \\ 
            &= (1-{\zeta}_k)\left(\frac{2}{\zeta_k} -1\right)\epsilon^2 \left\{\frac{1 - \left(1-\frac{\zeta_k}{2}\right)^{t}}{1-\left(1-\frac{\zeta_k}{2}\right)} \right\} \leq \frac{4(1-{\zeta}_k)\epsilon^2}{{\zeta}_k^2},
        \end{align}
        where $(a)$ follows from $\boldsymbol{\Delta_k}^{(0)} = \boldsymbol{0}_N$.
        This completes the proof.

	\section{Proof of Theorem~1}\label{Apdx:Thm1}	
	Suppose a fixed learning rate $\eta$ (i.e., $\eta^{(t)} = \eta$, $\forall t$). 
        From Step 6 and Step 10 of {\bf Procedure~\ref{alg:FedSpar}}, the local model update reconstructed by device $k\in\mathcal{K}_t$ is expressed as
	\begin{align}
	    \hat{\boldsymbol g}_{k}^{(t)} &= {\sf LocalUpdate}({\boldsymbol w}^{(t)}, \mathcal{D}_k)  + {\boldsymbol \Delta}_k^{(t-1)} - {\boldsymbol \Delta}_k^{(t)}  \overset{(a)}{=} \nabla F_k^{(t,1)}\big({\boldsymbol w}^{(t)}\big) + {\boldsymbol \Delta}_k^{(t-1)} - {\boldsymbol \Delta}_k^{(t)},
	\end{align}
	where $(a)$ follows from \eqref{eq:local_model_update} with $E=1$ and ${\boldsymbol w}^{(t,1)} = {\boldsymbol w}^{(t)}$. 
	Similarly, from Step 19 of {\bf Procedure~\ref{alg:FedSpar}}, the global model update reconstructed at iteration $t$ with ${\rho}_k= 1/M$ is expressed as
         \begin{align}\label{eq:reconstructPS}
            {\boldsymbol g}_{\rm PS}^{(t)}& =
            \frac{1}{M}\sum_{k\in\mathcal{K}_t} \hat{\boldsymbol{g}}_{k}^{(t)} = 
            \frac{1}{M}\sum_{k\in\mathcal{K}_t}  \big\{\nabla F_k^{(t,1)}\big({\boldsymbol w}^{(t)}\big) +{\boldsymbol \Delta}_k^{(t-1)} - {\boldsymbol \Delta}_k^{(t)} \big\}.
	\end{align}
	Then, from Step 20 of {\bf Procedure~\ref{alg:FedSpar}}, the global parameter vector $\boldsymbol{w}^{(t+1)}$ at iteration $(t+1)$ with is expressed as
        \begin{align}\label{eq:parametervector}
	    {\boldsymbol w}^{(t+1)}  = {\boldsymbol w}^{(t)}  
     - \eta \bigg\{&\frac{1}{M}\sum_{k\in\mathcal{K}_t} \nabla F_k^{(t,1)}\big({\boldsymbol w}^{(t)}\big) +  \frac{1}{M}\sum_{k\in\mathcal{K}_t} \big({\boldsymbol \Delta}_k^{(t-1)} - {\boldsymbol \Delta}_k^{(t)} \big)\bigg\}. 
	\end{align}
	Let $\tilde{\boldsymbol w}^{(t)}$ be an {\em ideal} parameter vector at iteration $t$ obtained when there are no residual errors. 
	Then the ideal parameter vector is updated as 
	\begin{align}\label{eq:non-error-parametervector}
    	\tilde{\boldsymbol w}^{(t+1)} = \tilde{\boldsymbol w}^{(t)} -  \frac{\eta}{M}\sum_{k\in\mathcal{K}_t}\nabla F_k^{(t,1)}\big({\boldsymbol w}^{(t)}\big),
	\end{align}
	while $\tilde{\boldsymbol w}^{(1)} = {\boldsymbol w}^{(1)}$ by the definition. 
	The comparison between \eqref{eq:parametervector} and \eqref{eq:non-error-parametervector} implies that the ideal parameter vector at iteration $t$ is expressed as
	\begin{align}\label{eq:non-error-parametervector_vs_parametervector}
    	\tilde{\boldsymbol w}^{(t)} &= {\boldsymbol w}^{(t)} + \sum_{i=1}^{t-1} \eta\left\{ \frac{1}{M}\sum_{k\in\mathcal{K}_t} \big({\boldsymbol \Delta}_k^{(i-1)} - {\boldsymbol \Delta}_k^{(i)} \big)\right\} \overset{(a)}{=} {\boldsymbol w}^{(t)} -  \frac{\eta}{M}\sum_{k\in\mathcal{K}}{\boldsymbol \Delta}_k^{(t-1)},
	\end{align}
	where $(a)$ follows from ${\boldsymbol \Delta}_{k}^{(0)} = {\bf 0}_N$ and Step 13 of {\bf Procedure~\ref{alg:FedSpar}}.
	
	Under Assumptions 1--2, the improvement of the loss function at iteration $t$ satisfies 
    \begin{align} \label{eq:convergence1}
        &\mathbb{E}[F(\tilde{\boldsymbol w}^{(t+1)}) - F(\tilde{\boldsymbol w}^{(t)})]  
        \leq
        \mathbb{E}[\nabla F(\tilde{\boldsymbol w}^{(t)})^{\sf T}(\tilde{\boldsymbol w}^{(t+1)}-\tilde{\boldsymbol w}^{(t)})]
            +  \frac{\beta}{2}\mathbb{E}[\|\tilde{\boldsymbol w}^{(t+1)}-\tilde{\boldsymbol w}^{(t)}\|^2]
        \nonumber \\ &\overset{(a)}=
        - \mathbb{E}\Bigg[\nabla F(\tilde{\boldsymbol w}^{(t)})^{\sf T} \frac{\eta}{M}\sum_{k\in\mathcal{K}_t}\nabla F_k^{(t,1)}\big({\boldsymbol w}^{(t)}\big)\Bigg] +\frac{\eta^2 \beta}{2}\mathbb{E}\Bigg[\bigg\|\frac{1}{M}\sum_{k\in\mathcal{K}_t}\nabla F_k^{(t,1)}\big({\boldsymbol w}^{(t)}\big)\bigg\|^2\Bigg] 
        \nonumber \\ &\overset{(b)}\leq
        -{\eta}\mathbb{E}[\nabla F(\tilde{\boldsymbol w}^{(t)})^{\sf T}\nabla F\big({\boldsymbol w}^{(t)}\big)] +\frac{\eta^2 \beta}{2M^2}\mathbb{E}\bigg[\bigg\|\sum_{k\in\mathcal{K}_t}\nabla F_k^{(t,1)}\big({\boldsymbol w}^{(t)}\big)\bigg\|^2\bigg], 
    \end{align}
    where $(a)$ follows from \eqref{eq:non-error-parametervector}, and $(b)$ holds because 
    \begin{align}
    &\frac{1}{M}\sum_{k\in\mathcal{K}_t} \mathbb{E}\big[\nabla F_k^{(t,1)}({\boldsymbol w}^{(t)})\big] = \frac{1}{M}\sum_{k\in\mathcal{K}_t} \nabla F_k({\boldsymbol w}^{(t)}) 
    = \nabla F({\boldsymbol w}^{(t)}),
    \end{align}
    under Assumption 2. 
    For any real vectors ${\boldsymbol u}_1,\dots,{\boldsymbol u}_I\in\mathbb{R}^{{N}}$, we have $2{\boldsymbol u}_i^{\sf T} {\boldsymbol u}_j = \|{\boldsymbol u}_i\|^2 + \|{\boldsymbol u}_j\|^2 - \|{\boldsymbol u}_i - {\boldsymbol u}_j\|^2, \forall i,j \in \{1,\dots,I\}$ and $\|\sum_{i=1}^I{\boldsymbol u}_i\|^2 \leq I\sum_{i=1}^I\|{\boldsymbol u}_i\|^2$. 
    Applying these inequalities into \eqref{eq:convergence1} yields
    \begin{align}\label{eq:convergence2}
        &\mathbb{E}[F(\tilde{\boldsymbol w}^{(t+1)}) - F(\tilde{\boldsymbol w}^{(t)})]  
        \nonumber\\
        &\leq
        -\frac{\eta}{2}\Big\{\mathbb{E}[\|\nabla F(\tilde{\boldsymbol w}^{(t)})\|^2] + \mathbb{E}[\|\nabla F\big({\boldsymbol w}^{(t)}\big)\|^2] - \mathbb{E}[\|\nabla F(\tilde{\boldsymbol w}^{(t)}) - \nabla F\big({\boldsymbol w}^{(t)}\big)\|^2] \Big\} \nonumber \\
        &~~~+ 
        \frac{\eta^2 \beta}{2M}\sum_{k\in\mathcal{K}_t} \mathbb{E}\big[\big\|\nabla F_k^{(t,1)}\big({\boldsymbol w}^{(t)}\big)\big\|^2\big], 
        \nonumber \\ 
        &\overset{(a)}\leq
        -\frac{\eta}{2}\mathbb{E}[\|\nabla F\big({\boldsymbol w}^{(t)}\big)\|^2] + \frac{\eta}{2}\mathbb{E}[\|\nabla F(\tilde{\boldsymbol w}^{(t)}) - \nabla F\big({\boldsymbol w}^{(t)}\big)\|^2] + \frac{\eta^2 \beta \epsilon^2}{2}
        \nonumber \\ &\overset{(b)}\leq
        -\frac{\eta}{2}\mathbb{E}[\|\nabla F\big({\boldsymbol w}^{(t)}\big)\|^2] + \frac{\eta\beta^2}{2}\mathbb{E}[\|\tilde{\boldsymbol w}^{(t)} - {\boldsymbol w}^{(t)}\|^2] + \frac{\eta^2 \beta \epsilon^2}{2},
    \end{align}
    where $(a)$ follows from $-\frac{\eta}{2}\mathbb{E}[\|\nabla F\big(\tilde{\boldsymbol w}^{(t)}\big)\|^2] \leq 0$ and  $\mathbb{E}\big[\big\|\nabla F_k^{(t,e)}({\boldsymbol w})\big\|^2\big] \leq \epsilon^2$ under Assumption 2, and $(b)$ holds because the loss function $F({\boldsymbol w})$ is $\beta$-smooth under Assumption 1. 
    From \eqref{eq:non-error-parametervector_vs_parametervector}, the second expectation term in the right-hand-side (RHS) of \eqref{eq:convergence2} can be  rewritten as
    \begin{align}\label{eq:expect_convergence2}
        \mathbb{E}[\|\tilde{\boldsymbol w}^{(t)} - {\boldsymbol w}^{(t)}\|^2] 
        = \mathbb{E}\Bigg[\bigg\| -\frac{\eta}{M}\sum_{k\in\mathcal{K}}{\boldsymbol \Delta}_k^{(t-1)}\bigg\|^2\Bigg]  
        \overset{(a)}\leq \frac{\eta^2K}{M^2}\sum_{k\in\mathcal{K}} \mathbb{E}\big[\big\| {\boldsymbol \Delta}_k^{(t-1)} \big\|^2\big]
        \overset{(b)}\leq \frac{\eta^2K^2}{M^2} \max_k \delta_k^2,
    \end{align}
    where $(a)$ follows from $\|\sum_{i=1}^I{\boldsymbol u}_i\|^2 \leq {I}\sum_{i=1}^I\|{\boldsymbol u}_i\|^2$, and $(b)$ follows from Lemma 1. 
    Applying the above inequality and $\eta=\frac{2}{\sqrt{T}}$ into the RHS of \eqref{eq:convergence2} yields 
    \begin{align}\label{eq:convergence3}
       &\mathbb{E}[F(\tilde{\boldsymbol w}^{(t+1)}) - F(\tilde{\boldsymbol w}^{(t)})]  \leq
        -\frac{1}{\sqrt{T}}\mathbb{E}[\|\nabla F\big({\boldsymbol w}^{(t)}\big)\|^2]
        + \frac{4K^2}{M^2 T\sqrt{T}}\beta^2\max_k\delta_k^2 + \frac{2}{T}\beta \epsilon^2.
    \end{align}
    Then summing both sides of \eqref{eq:convergence3} over the iterations yields 
    \begin{align}\label{eq:convergence4}
        &\sum_{t=1}^T\mathbb{E}[F(\tilde{\boldsymbol w}^{(t+1)}) - F(\tilde{\boldsymbol w}^{(t)})] = \mathbb{E}[F(\tilde{\boldsymbol w}^{(T+1)}) - F(\tilde{\boldsymbol w}^{(1)})]
        \nonumber \\ &\leq
        -\frac{1}{\sqrt{T}} \sum_{t=1}^T\mathbb{E}[\|\nabla F\big({\boldsymbol w}^{(t)}\big)\|^2] + \frac{4K^2}{M^2\sqrt{T}}\beta^2\max_k\delta_k^2 + {2 \beta \epsilon^2}.
    \end{align}
    The above inequality can be rewritten as
    \begin{align}\label{eq:convergence5}
        \mathbb{E}\Bigg[\frac{1}{{T}}\sum_{t=1}^T\|\nabla F\big({\boldsymbol w}^{(t)}\big)\|^2\Bigg]
        &\overset{(a)}\leq
        \frac{1}{\sqrt{T}}\big\{F({\boldsymbol w}^{(1)}) - F({\boldsymbol w}^{\star})\big\} + \frac{4K^2}{M^2T}\beta^2\max_k\delta_k^2 + \frac{2}{\sqrt{T}} \beta \epsilon^2
        \nonumber \\ &\leq
        \frac{1}{\sqrt{T}}\left\{F({\boldsymbol w}^{(1)}) - F({\boldsymbol w}^{\star}) + 2\beta \epsilon^2 
        + \frac{4K^2}{M^2\sqrt{T}}\beta^2\max_k\delta_k^2\right\},
    \end{align}
    where $(a)$ follows from $\tilde{\boldsymbol w}^{(1)} = {\boldsymbol w}^{(1)}$ and $F(\boldsymbol{w})\geq F(\boldsymbol{w}^{\star})$, $\forall \boldsymbol{w}\in \mathbb{R}^{{N}}$. This completes the proof.


\begin{thebibliography}{1}
     
        \bibitem{Konecny:2015} J. Kone\u{c}n\'{y}, H. B. McMahan, and D. Ramage, ``Federated optimization: Distributed optimization beyond the datacenter,'' in {\em Adv. Neural Inf. Process. Syst. Workshop Optim. Mach. Learn.,} Montreal, QC, Canada, Dec. 2015, pp. 1--5.
    
    	\bibitem{Mcmahan:2017} H. B. McMahan, E. Moore, D. Ramage, S. Hampson, and B. A. Arcas, ``Communication-efficient learning of deep networks from decentralized data,'' in {\em Proc. Int. Conf. Artif. Intell. Statist. (AISTATS),} Fort Lauderdale, FL, USA, Apr. 2017, pp. 1273--1282.
          
        \bibitem{Konecny:2017} J. Kone\u{c}n\'{y}, H. B. McMahan, F. X. Yu, P. Richt\'{a}rik, A. T. Suresh, and D. Bacon, 
        ``Federated learning: Strategies for improving communication efficiency,'' in {\em Adv. Neural Inf. Process. Syst. Workshop Private Multi-Party Mach. Learn.,} Barcelona, Spain, Dec. 2016, pp. 1--5.
            
    	\bibitem{Niknam:2020} S. Niknam, H. S. Dhillon, and J. H. Reed, ``Federated learning for wireless communications: Motivation, opportunities, and challenges,'' {\em IEEE Commun. Mag.,} vol. 58, no. 6, pp. 46--51, Jun. 2020.
        
    	\bibitem{Zhu:2020} G. Zhu, D. Liu, Y. Du, C. You, J. Zhang, and K. Huang, ``Toward an intelligent edge: Wireless communication meets machine learning,'' 
          {\em IEEE Commun. Mag.,} vol. 58, no. 1, pp. 19--25, Jan. 2020.
        
    	\bibitem{Gunduz:2020} D. G\"{u}nd\"{u}z, D. B. Kurka, M. Jankowski, M. M. Amiri, E. Ozfatura, and S. Sreekumar, ``Communicate to learn at the edge,''
          {\em IEEE Commun. Mag.,} vol. 58, no. 12, pp. 14--19, Dec. 2020.
    
        

    
    	\bibitem{ResNet} K. He, X. Zhang, S. Ren, and J. Sun, ``Deep residual learning for image recognition,'' in {\em Proc. Comput. Vis. Pattern Recog. (CVPR),} Las Vegas, NV, USA, Jun. 2016, pp. 770--778.	
    	

	
        
        \bibitem{Nguyen:21} H. T. Nguyen, V. Sehwag, S. Hosseinalipour, C. G. Brinton, M. Chiang, and H. V. Poor, ``Fast-convergent federated learning,'' {\em IEEE J. Sel. Areas Commun.,} vol. 39, no. 1, pp. 201--218, Jan. 2021.
        
    	\bibitem{Mingzhe:21} M. Chen, H. V. Poor, W. Saad, and S. Cui, ``Convergence time optimization for federated learning over wireless networks,'' {\em IEEE Trans. Wireless Commun.,} vol. 20, no. 4, pp. 2457--2471, Apr. 2021.
    	
    	\bibitem{Hao:22} H. Chen, S. Huang, D. Zhang, M. Xiao, M. Skoglund, and H. V. Poor, 
           ``Federated learning over wireless IoT networks with optimized communication and resources,'' 
           {\em IEEE Internet Things J.,} vol. 9, no. 17, pp. 16592--16605, Sep. 2022. 
           
    	\bibitem{Jeon:2021} Y.-S. Jeon, M. M. Amiri, J. Li, and H. V. Poor, ``A compressive sensing approach for federated learning over massive MIMO communication systems,''
          {\em IEEE Trans. Wireless Commun.,} vol. 20, no. 3, pp. 1990--2004, Mar. 2021.
          
        
        
                   
    	\bibitem{SignSGD} J. Bernstein, Y.-X. Wang, K. Azizzadenesheli, and A. Anandkumar, ``SignSGD: Compressed optimisation for non-convex problems,'' 
            in {\em Proc. Int. Conf. Mach. Learn. (ICML),} Stockholmsm\"{a}ssan, Stockholm, Sweden, Jul. 2018, pp. 560--569.
        
        \bibitem{QSGD} D. Alistarh, D. Grubic, J. Li, R. Tomioka, and M. Vojnovic, ``QSGD: Communication-efficient SGD via gradient quantization and encoding,'' in {\em Adv. Neural Inf. Process. Syst. (NIPS),} Long Beach, CA, USA, Dec. 2017, pp. 1707--1718.
        
        \bibitem{VeQ:1} N. Shlezinger, M. Chen, Y. C. Eldar, H. V. Poor, and S. Cui, ``UVeQFed: Universal vector quantization for federated learning,''
            {\em IEEE Trans. Signal Process.,} vol. 69, pp. 500--514, Dec. 2020.
        
        \bibitem{VeQ:2} M. Chen, N. Shlezinger, H. V. Poor, Y. C. Eldar, and S. Cui, ``Communication efficient federated learning,'' in {\em Proc. Nat. Acad. Sci.,} vol. 118, no. 17, Apr. 2021.
    
        \bibitem{VeQ:3} Y. Du, S. Yang, and K. Huang, ``High-dimensional stochastic gradient quantization for communication-efficient edge learning,''
          {\em IEEE Trans. Signal Process.,} vol. 68, pp. 2128--2142, Mar. 2020.
        
        \bibitem{Zheng:2020} S. Zheng, C. Shen, and X. Chen, ``Design and analysis of uplink and downlink communications for federated learning,'' {\em IEEE J. Sel. Areas Commun.,} vol. 39, no. 7, pp. 2150--2167, Jul. 2021.
        
        \bibitem{QCS:1} J. Wangni, J. Wang, J. Liu, and T. Zhang, ``Gradient sparsification for communication-efficient distributed optimization,''
          in {\em Adv. Neural Inf. Process. Syst. (NeurIPS),} Montreal, QC, Canada, Dec. 2018, pp. 1306--1316.
        
        \bibitem{QCS:2} Y. Lin, S. Han, H. Mao, Y. Wang, and W. J. Dally, ``Deep gradient compression: Reducing the communication bandwidth for distributed training,''
          in {\em Proc. Int. Conf. Learn. Represent. (ICLR),} Vancouver, BC, Canada, Apr. 2018, pp. 1--14.
          
        \bibitem{Jeon:2022} Y.-S. Jeon, M. M. Amiri, and N. Lee, ``Communication-efficient federated learning over MIMO multiple access channels,'' {\em IEEE Trans.  Commun.,} vol. 70, no. 10, pp. 6547--6562, Oct. 2022.  
        
        \bibitem{ATOMO} H. Wang, S. Sievert, S. Liu, Z. Charles, D. Papailiopoulos, and S. Wright, ``Atomo: Communication-efficient learning via atomic sparsification,'' in {\em Adv. Neural Inf. Process. Syst. (NeurIPS),} Montreal, QC, Canada, Dec. 2018, pp. 9872--9883.   
        
        \bibitem{PowerSGD} T. Vogels, S. P. Karimireddy, and M. Jaggi, ``PowerSGD: Practical low-rank gradient compression for distributed optimization,'' in {\em Adv. Neural Inf. Process. Syst. (NeurIPS),} Vancouver, BC, Canada, Dec. 2019, pp. 14269--14278.   
               
           

        
        
        \bibitem{SQCS:1} X. Fan, Y. Wang, Y. Huo, and Z. Tian, "1-bit compressive sensing for efficient federated learning over the air," {\em IEEE Trans. Wireless Commun.,} vol. 22, no. 3, pp. 2139--2155, Mar. 2023.
        
        \bibitem{SQCS:2} C. Li, G. Li, and P. K. Varshney, 
           ``Communication-efficient federated learning based on compressed sensing,'' 
           {\em IEEE Internet Things J.,} vol. 8, no. 20, pp. 15531--15541, Oct. 2021. 
     
        \bibitem{FedQCS} Y. Oh, N. Lee, Y.-S. Jeon, and H. V. Poor, ``Communication-efficient federated learning via quantized compressed sensing,'' {\em IEEE Trans. Wireless Commun.,} vol. 22, no. 2, pp. 1087--1100, Feb. 2023.
        
        \bibitem{FedVQCS} Y. Oh, Y.-S. Jeon, M. Chen, and W. Saad, ``FedVQCS: Federated learning via vector quantized compressed sensing,'' {\em IEEE Trans. Wireless Commun.,} to be published.


        \bibitem{D-DSGD} M. M. Amiri and D. G\"{u}nd\"{u}z, ``Machine learning at the wireless edge: Distributed stochastic gradient descent over-the-air,'' 
          {\em IEEE Trans. Signal Process.,} vol. 68, pp. 2155--2169, Mar. 2020.
        
        \bibitem{F. Sattler:2019} F. Sattler, S. Wiedemann, K.-R. M\"{u}ller, and W. Samek, ``Sparse binary compression: Towards distributed deep learning with minimal communication,” in {\em Proc. IEEE Int. Joint Conf. Neural Netw. (IJCNN),} Budapest, Hungary, Jul. 2019, pp. 1--8.

        \bibitem{TCS} E. Ozfatura, K. Ozfatura, and D. G\"{u}nd\"{u}z, ``Time-correlated sparsification for communication-efficient federated learning,'' in {\em Proc. Int. Symp. Inf. Theory (ISIT),} Melbourne, Australia, Jul. 2021, pp. 461--466.  
    
        \bibitem{AmiriDownlink} M. M. Amiri, D. G\"{u}nd\"{u}z, S. R. Kulkarni, and H. V. Poor, ``Convergence of federated learning over a noisy downlink,'' {\em IEEE Trans. Wireless Commun.,} vol. 21, no. 3, pp. 1422--1437, Mar. 2022.
        

        \bibitem{J. J. Bussgang:1952} J. J. Bussgang, ``Crosscorrelation functions of amplitude-distorted Gaussian signals,'' MIT Res. Lab. Electron., Cambridge, MA, USA, Rep. 216, Mar. 1952.

        \bibitem{MNIST} Y. Lecun, L. Bottou, Y. Bengio, and P. Haffner, ``Gradient-based learning applied to document recognition,'' {\em Proc. IEEE} vol. 86, no. 11, pp. 2278--2324, Nov. 1998.
    

        \bibitem{CIFAR-10} A. Krizhevsky, ``Learning multiple layers of features from tiny images,'' M.S. thesis, Univ. Toronto, Toronto, ON, Canada, 2009.

        \bibitem{S. P. Lloyd:1982} S. P. Lloyd, ``Least squares quantization in PCM,'' 
            {\em IEEE Trans. Inf. Theory,} vol. 28, no. 2, pp. 129--137, Mar. 1982.
     
        \bibitem{Lexicographic} M. C. Er, ``Lexicographic ordering, ranking and unranking of combinations,'' {\em Int. J. Comput. Math.,} vol. 17, no. 3--4, pp. 277--283, Nov. 1985.
        
        \bibitem{Convergence} H. Yu, S. Yang, and S. Zhu, ``Parallel restarted SGD with faster convergence and less communication: Demystifying why model averaging works for deep learning,'' {\em AAAI,} vol. 33, no. 01, pp. 5693--5700, Jul. 2019.

        \bibitem{FedCOM} F. Haddadpour, M. M. Kamani, A. Mokhtari, and M. Mahdavi, ``Federated learning with compression: Unified analysis and sharp guarantees,'' in {\em Proc. Int. Conf. Artif. Intell. Statist. (AISTATS),} Apr. 2021, pp. 2350--2358.
        
        
        \bibitem{Adam} D. P. Kingma and J. Ba, ``Adam: A method for stochastic optimization,'' in {\em Proc. Int. Conf. Learn. Represent. (ICLR),} San Diego, CA, USA, May 2015, pp. 1--13.
        
        
        

        \bibitem{ShuffleNet} X. Zhang, X. Zhou, M. Lin, and J. Sun, ``ShuffleNet: An extremely efficient convolutional neural network for mobile devices,'' in {\em Proc. Comput. Vis. Pattern Recog. (CVPR),} Salt Lake City, UT, USA, Jun. 2018, pp. 6848--6856.

        \bibitem{PathLoss} V. Erceg {\em et al}., ``An empirically based path loss model for wireless channels in suburban environments,'' {\em IEEE J. Sel. Areas Commun.,} vol. 17, no. 7, pp. 1250--1211, Jul. 1999.


        \bibitem{ErrorAccum} S. P. Karimireddy, Q. Rebjock, S. Stich, and M. Jaggi, ``Error feedback fixes signSGD and other gradient compression schemes,'' in {\em Proc. Int. Conf. Mach. Learn. (ICML),} Long Beach, CA, USA, Jun. 2019, pp. 3252--3261.
        
    \end{thebibliography}
\end{document}